\documentclass{iopart}
\usepackage{iopams,graphicx}
\usepackage{graphicx}

\def\be{\begin{equation}}
\def\ee{\end{equation}}
\def\bea{\begin{eqnarray}}
\def\eea{\end{eqnarray}}

\begin{document}


\title{Constraints on the Inflationary Expansion from Three Year WMAP, small scale CMB anisotropies
and Large Scale Structure Data Sets}
  
\author{F Finelli\dag\ddag\S, M Rianna\dag, N Mandolesi\dag}
\address{\dag INAF/IASF-BO, 
Istituto di Astrofisica Spaziale e Fisica 
Cosmica di Bologna \\
via Gobetti 101, I-40129 Bologna - Italy}
\address{\ddag INAF/OAB, Osservatorio Astronomico di Bologna, 
via Ranzani 1, I-40127 Bologna -
Italy}
\address{\S INFN, Sezione di Bologna,
Via Irnerio 46, I-40126 Bologna, Italy}

\eads{\mailto{finelli@iasfbo.inaf.it},
\mailto{rianna@iasfbo.inaf.it}, \mailto{mandolesi@iasfbo.inaf.it}}

\date{\today} 

\begin{abstract} 
We present the constraints on inflationary parameters in a flat 
$\Lambda$CDM universe obtained by WMAP 
three year data release, plus smaller scale CMB and two LSS data sets, 2dF 
and SDSS (treated separately). We 
use a Markov Chain Monte Carlo (MCMC) technique combined with an analytic 
description of the inflationary spectra in terms of the horizon 
flow functions (HFF).  
By imposing a consistency condition for the tensor-to-scalar ratio,
we study the constraints both on single field standard 
inflation and on inflation with the violation of the null energy condition, 
which leads to a blue spectrum for gravitational waves. 
For standard inflation, the constraint on the tensor-to-scalar
ratio we obtain from CMB data and 2dF05 is: $r_{0.01} < 0.26$ 
at 2 $\sigma$ cl.
Without the consistency condition between the tensor-to-scalar ratio and the tensor slope, 
the constraints on the tensor amplitude is not significantly changed, 
but the constraints on the HFFs are significantly relaxed. 
We then show that when the third HFF $\epsilon_3$ is allowed 
to be non-zero and to be of order unity, 
a large negative (at $2 \sigma$) 
value for the running of the scalar spectral 
index in standard inflation is found in any set of data we consider.
\end{abstract}

\pacs{CMBR theory, inflation}

\maketitle

\section*{Introduction}

Inflation is the most promising paradigm for understanding the initial
conditions for cosmic structure formation and the pattern of anisotropies
in temperature and polarization of the cosmic microwave background (CMB).
Its predictions for the simplest case of a single scalar field model with a
nearly scale-invariant spectrum of curvature and tensor perturbations are in
good agreement with most of the observational data.

Among observations, data on CMB anisotropies have been the most
selective for inflation, starting from COBE-DMR \cite{cobe} to the NASA 
mission presently in operation WMAP \cite{wmap}. The 
WMAP first-year data \cite{hinshaw_1,kogut} 
complemented with smaller scale data started 
to discriminate among the inflationary models 
and to constrain the possibility that the spectra may have not a 
pure power-law shape \cite{peiris,leachliddle}.

In this paper we compare inflationary predictions with WMAP three year 
data release \cite{hinshaw,page,spergel}
plus smaller scale CMB data (which we shall denote collectively as 
CMBsmall), such as CBI \cite{readhead}, ACBAR 
\cite{kuo}, VSA \cite{dickinson} and B2K \cite{jones,piacentini,montroy}.  
We shall also use separately two LSS data sets, SDSS \cite{tegmark} and 2dF 
\cite{percival,cole}. 

We compare inflationary predictions with observations, 
by adopting a parametrization of the 
primordial power spectra (PS henceforth) of curvature and tensor 
perturbations as:
\begin{eqnarray}
P_S (k) &=& A_S \, e^{(n_S - 1) \ln (k/k_*) + \alpha_S \ln^2(k/k_*)/2}
\,, \nonumber \\
P_T (k) &=& A_T \, e^{n_T \ln (k/k_*) + \alpha_T \ln^2(k/k_*)/2}
\label{i_spectra}
\end{eqnarray}
where $i={\rm S \,, T}$ stands for scalar and tensor, respectively, $k_*$
is the pivot scale. We shall use $A_i, n_i, \alpha_i$ in terms of 
of the Hubble parameter $H$ and
the {\em horizon flow functions}
$\epsilon_i$ (HFF henceforth) evaluated at the pivot scale $k_*$. 
At the present state of the art, the PS of primordial perturbations 
in most of the single field inflationary models depend only 
on the physics at the Hubble crossing since adiabaticity on large scale 
is preserved.

The HFFs are defined as $\epsilon_1 = - \dot
H/H^2$ and $\epsilon_{i + 1} \equiv \dot \epsilon_i/(H \epsilon_i) = 
(d \epsilon_i/d \, N)/\epsilon_i$ with $i
\ge 1$ and $N$ the number fo e-folds ($d N = H dt$) 
\cite{HFF_ref}. For a Klein-Gordon scalar field $\phi$, $H$ and 
$\epsilon_i$ are related to the potential and its derivatives as:
\begin{eqnarray}
V &=& 3 \, M_{\rm Pl}^2 \, H^2 \left(1 - 
\frac{\epsilon_1}{3}\right), \nonumber \\
V_\phi &=& - 3 \, \sqrt{2} \, M_{\rm Pl} \, H^2 \, 
\epsilon_1^{1/2} \, \left(1 - \frac{\epsilon_1}{3} + 
\frac{\epsilon_2}{6}\right), \nonumber \\
\frac{V_{\phi \phi}}{3 H^2} &=& 2 \epsilon_1 - \frac{\epsilon_2}{2}
- \frac{2\epsilon_1^2}{3} + \frac{5\epsilon_1 \epsilon_2}{6} 
-\frac{\epsilon_2^2}{12} - \frac{\epsilon_2 \epsilon_3}{6} \,,
\label{potential_eps}
\end{eqnarray}
where $M_{\rm Pl} = (8 \pi G)^{-1/2} \simeq 2.4 \times 10^{18}$ Gev is 
the reduced Planck mass. By comparing constraints 
on $H \,, \epsilon_i$ derived by observations, the shape of the inflationary 
potential can be constrained through Eqs. (\ref{potential_eps}).

We use the most advanced analytic descriptions of inflationary 
spectra which relate $(A_i, n_i, \alpha_i)$ to $(H, \epsilon_j)$. 
These expressions are obtained by the Green's function method (henceforth GFM) 
\cite{GS,LLMS} and by the method of comparison equations (henceforth MCE)
\cite{CFKLV}. These two methods provide a description of cosmological 
perturbations spectra up to second order in the HFF, 
allowing a good accuracy also in inflationary models where the HFF are not 
so small or constant in time, i.e. the slow-roll condition is 
not well satisfied \cite{LLMS,CFKLV,makarov}. Note that the GFM and MCE  
methods predict the same PSs at ${\cal O} (\epsilon_j)$, but differences 
of percent order arise in the coefficients of ${\cal O} (\epsilon_j^2)$ 
terms \cite{CFKLV}.


In this paper we address several points of current interest in 
comparing inflationary preditions with observations. 
At first order in HFFs we update the analysis of Leach and Liddle 
\cite{leachliddle} to WMAP3 \cite{hinshaw,page,spergel} 
and to 2dF05 \cite{cole}. At second order in HFFs, 
we first show how the large running found by the WMAP team 
\cite{spergel_1,peiris,spergel} 
is recovered by using the PS parametrization in terms of the HFFs. 
We also give a model-independent constraint on the tensor amplitude by 
considering the tensor-to-scalar ratio as a free parameter and not given  
in terms of the tensor spectral index. We then study the constraints on 
inflationary models which violate the null energy condition (NEC) and 
predict a blue spectrum for gravitational waves \cite{BFM}.


Our paper is organized as follows. In section II we present our methodology 
and in section III the 
constraints on cosmological parameters by restricting the parametrization 
for the spectra at first order in HFFs for standard inflationary models.  
In section III we relax the assumption that the inflaton is a Klein-Gordon 
(KG henceforth)
scalar field and we leave the tensor-to-scalar ratio as a free parameter. 
In section IV we compare the constraints on inflationary models 
with a blue spectrum for gravitational waves with standard inflationary 
models. In section V we present the results obtained allowing running(s) of the 
spectral indices in standard inflationary models, i.e. taking into account 
spectra to second order in HFFs. In section VI we conclude.

\section{Methodology}

We use the publicly available Markov Chain Monte Carlo (MCMC) package 
\cite{LB} which makes use of the Einstein-Botzmann code CAMB 
\cite{camb} for the 
computation of theoretical CMB anisotropies PSs and 
transfer functions. We restrict ourselves to flat spatial sections and 
to three species of massless neutrinos. 
We use as initial conditions for cosmological 
fluctuations the growing adiabatic mode. 
We include the CMBsmall data set as by default
implemented in COSMOMC. Note that small scale data
by the same experiments have been used by the WMAP team
in Ref. [9], although there may be additional cross-correlations
between the data sets which we consider and WMAP3 data.
We restrict our analysis to linear 
transfer functions and therefore the LSS data are automatically 
truncated by the MCMC to $k_{\rm cut}/h \simeq 0.15$ ($0.2$) for 2dF (SDSS). 
The convergence diagnostic on runs with multiple chains is the R statistic 
by Gelman and Rubin implemented in the MCMC. 

We use the $\ln(k/k_*)$ expansion of the logarithm of the 
parametrizations in Eqs. (\ref{i_spectra}), in order to avoid that the 
spectrum becomes negative for some values of the inflationary 
parameters explored by MCMC \cite{LLMS}. 
According to Eqs. (\ref{i_spectra}), the $\ln(k/k_*)$ 
expansion of the power spectra is:
\begin{equation}
\label{plex}
\ln {P (k)\over P_0(k_*)} = b_0 + b_1 \ln \left(k\over 
k_*\right)
+ \frac{b_2}{2} \ln^2\left(k\over k_*\right) + \dots .
\end{equation}
and we use both the GFM \cite{GS} and MCE \cite{LLMS} 
to relate the coefficients $b$s to the HFFs. 
The coefficients for the scalar spectrum are:  
\begin{eqnarray}
\label{eqn:bs0}
b_{{\rm S}0} &=&
- 2\left(C + 1\right)\epsilon_1 - C \epsilon_2
+ \left(- 2C + {\textstyle\frac{\pi^2}{2}} - 7\right)
 \epsilon_1^2 \nonumber\\
& & + \left({\textstyle\frac{\pi^2}{8}} - 1\right)
\epsilon_2^2 +
\left(- X^2 - 3X + {\textstyle\frac{7\pi^2}{12}} - 7 + 2 \Delta_{{\rm S} 0}
\right)
\epsilon_1\epsilon_2
\nonumber \\
 & & + \left(-{\textstyle\frac 12}X^2 + 
{\textstyle\frac{\pi^2}{24}} + \Delta_{{\rm S} 0} \right)
\frac{d \epsilon_2}{d N} \, \\
\label{eqn:bs1}
b_{{\rm S}1} &\equiv&  n_S - 1 = - 2 \epsilon_1 - \epsilon_2 - 2 
\epsilon_1^2 -\left(2\,X+3\right)\,\epsilon_1\,\epsilon_2
-X\, \frac{d \epsilon_2}{d N} \, \\
b_{{\rm S}2} &\equiv& \alpha_S = - 2 \epsilon_1 \epsilon_2 - 
\frac{d \epsilon_2}{d N}
\label{eqn:bs2},
\end{eqnarray}
and for tensors are:
\begin{eqnarray}
b_{{\rm T}0} &=&
 - 2\left(C + 1\right)\epsilon_1
 + \left(- 2C + {\textstyle\frac{\pi^2}{2}} - 7\right)
 \epsilon_1^2 \nonumber\\
 & & + \left(-X^2 - 2X + {\textstyle\frac{\pi^2}{12}} - 2 + 
\Delta_{{\rm T} 0} \right)
 \epsilon_1\epsilon_2 , 
\label{eqn:bt0}
\\
b_{{\rm T}1} &\equiv&  n_T = - 2\epsilon_1 - 2\epsilon_1^2
-2\,\left(X+1\right)\,\epsilon_1\,\epsilon_2 , \\
\label{eqn:bt1}
b_{{\rm T}2} &\equiv&  \alpha_T = - 2\epsilon_1\epsilon_2,
\label{eqn:bt2}
\end{eqnarray}
where $C \equiv \ln 2+\gamma_{\rm E}-2\approx-0.7296$ ($\gamma_{\rm E}$ is
the Euler-Mascheroni constant) and 
$d \epsilon_2 / d N = \epsilon_2 \epsilon_3$. 
At second order the next-to-leading 
coefficients depend on the method of 
approximation used to study cosmological perturbations during inflation: 
$X=C$ and $\Delta_{{\rm S} 0} = \Delta_{{\rm T} 0} = 0$ are obtained 
within the GFM \cite{GS,LLMS}; $X= D \equiv \frac{1}{3}-\ln 
3\approx-0.7652$, $\Delta_{{\rm S} 0} = (D-C) (D + \ln \, 2) -1/18$, 
$\Delta_{{\rm T} 0} = 2D(D-C) -1/9$, 
within the MCE \cite{CFKLV}. We therefore implement the above expansion in a 
modified 
version of the inflationary routines developed by Samuel Leach and public at 
http://astronomy.sussex.ac.uk/$\sim$sleach/inflation/camb\_inflation.html.

As pivot scale we choose $k_*=0.01 \, {\rm Mpc}^{-1}$ as in
\cite{leachliddle} (see however \cite{LPLM}); this choice differs from 
those by the WMAP team
($k_*=0.002 \, {\rm Mpc}^{-1}$
\cite{peiris,spergel} and $k_*=0.05 \, {\rm Mpc}^{-1}$ \cite{spergel_1})
and is better placed at the center of the range
of wavenumbers probed by cosmological perturbations studied in the 
linear regime.
Although the CMB spectrum includes at least a convolution of the power-spectrm with spherical Bessel functions,  
an effective pivot $\ell_*$ corresponding to $k_*$ can be defined \cite{huthesis,LLMS}:
\be
k_*= \frac{3}{2} \frac{h \sqrt{\Omega_{\rm m}}}{1+0.084
\ln\Omega_{\rm m}}\ell_* \times 10^3 \,{\rm Mpc}^{-1} \,.
\ee
Since $\ell_* \simeq 145$ corresponding to $k_*=0.01 \, {\rm Mpc}^{-1}$ 
does not represent an interesting reference scale 
to measure the tensor-to-scalar ratio, we shall also quote in our results 
both $r_{k_*}$ and the ratio of tensor to scalar contribution to $C_\ell$ and $l=10$, 
denoted by $R_{10}$ ($\equiv C_{10}^T/C_{10}^S$).

As parameter describing reionization we shall vary the redshift $z_{\rm re}$, which is related to the 
optical depth $\tau$ by an integration over the redshift $z$ \cite{LB}
\be
\tau = \sigma_T \int_0^{z_{\rm re}} d z \frac{n_e (z)}{H(z) (1+z)^2} \,,
\ee
with $\sigma_T$ as the Thompson cross section and $n_e$ is the electron density number. 
As for the tensor-to-scalar ratios $r$ and $R_{10}$, we also give the constraints on $\tau$ as a derived parameter.

\section{First Order HFF Results}

In this section we explore the constraints for a model in which the 
inflaton is described by a Klein-Gordon scalar field 
and the spectra in Eqs. (\ref{plex}) are taken to first order in HFFs (i.e. 
$b_{{\rm S} 2} = b_{{\rm T} 2} = 0$).
When the inflaton is described by a Klein-Gordon scalar field the 
tensor-to-scalar ratio is not a free parameter, 
but is given by the so-called consistency condition which is:
\be
\frac{P_{0 \, T}}{P_{0 \, S}} = 16 \epsilon_1\,.
\label{ttos}
\ee
In this case we vary $7$ free parameters:
\be
\{\Omega_b \, h^2, \Omega_{cdm} h^2, H_0, 10^{10} A_{\rm S}, z_{re},
\epsilon_1,\epsilon_2 \} \,,
\label{string}
\ee
where $\Omega_b \,, \Omega_{cdm}$ are baryon and cold dark matter densities, $H_0$ is the present Hubble 
parameter, $h=H_0/(100 \, {\rm km \, s}^{-1} \, {\rm Mpc}^{-1})$, $A_{\rm S} 
(k_*)$ is the amplitude of super-Hubble curvature 
perturbation deep in the radiation era, $\epsilon_1 \,, \epsilon_2$ are the first 
two HFFs. The prior on these parameters are: $
0.005<\Omega_b h^2<0.1$, $0.01<\Omega_{cdm}<0.99$, 
$40<H_0/{\rm km s}^{-1} {\rm Mpc}^{-1}<100$, $4<z_{re}<30$ as in 
\cite{leachliddle} , 
$10^{-4}<\epsilon_1< 0.1$; $-0.25<\epsilon_2<0.2$ 


The one dimensional posterior marginalized probability distributions 
obtained from a Markov chain of (at least) $10^5$ elements 
(which uses the covariance matrix of a preliminary chain) 
for sampled and derived parameters are shown in Fig. 1,2 and in Table 1. 
We compare the results obtained by using the first year and 
the three years of WMAP data (plus CMBsmall and 2dF02 \cite{percival}), as 
well as the results of the three years
of WMAP data and CMBsmall with different LSS data sets (2dF02 
\cite{percival}, SDSS \cite{tegmark} and 2dF05 \cite{cole}). 
Different estimates of cosmological parameters obtained with different 
LSS data sets are very known (see Table 6 of \cite{spergel}) and may be due 
to different selection of the SDSS and 2dF surveys: this is the reason 
why we do not combine different LSS data sets.  

One of the main results of WMAP3, i.e. a better estimate 
(and different from WMAP1) of 
reionization, has changed the value of the scalar 
spectral index $n_S$. By combining with CMBsmall plus 2dF02, 
WMAP3 fully marginalized $2 \sigma$ value is $n_S=0.960 
\pm^{0.033}_{0.032}$ (compared to the WMAP1 correspondent value $n_S=0.979
\pm^{0.056}_{0.035}$).
As is clear from the right panel of Figs. 3,4, without marginalizing on the 
tensor-to-scalar ratio, a pure Harrison-Zeldovich 
spectrum for scalar perturbations with no tensors ($n_s=1 \,, r=0$) is disfavoured at more 
than $2 \sigma$ by the combination of CMB data with 2dF02 or SDSS, but is 
within $2 \sigma$ for CMB data plus 2dF05. 
Another interesting result we find is that inflationary models 
predicting $n_S = 1$ ($\epsilon_2 = -2 \epsilon_1$) {\em and} tensors 
are also disfavoured at $2 \sigma$ by the same data sets for which the HZ 
spectrum is disfavoured with the same cl (again, this is not true for CMB 
data plus 2dF05). As shown by Fig. 4, this effect is due to the inclusion 
of small scale CMB datasets 
\cite{readhead,kuo,dickinson,jones,piacentini,montroy}. 
Note from Fig. 4 how our results 
on WMAP3 + SDSS agree with previous analysis with the same data sets 
\cite{KKMR,PE}. 
By comparing our results to Fig. 1 of \cite{dominik} 
where WMAP1 results were 
reported, we see that models with a natural exit from inflation 
(the region $\epsilon_2 > 0$) are now preferred by WMAP3. 

\begin{figure}
\begin{tabular}{c}
\includegraphics[scale=0.4]{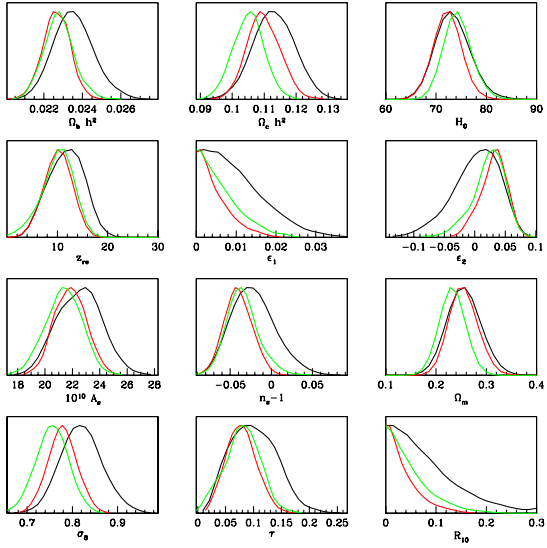}
\end{tabular}
\caption{One dimensional marginalized probabilities for 
cosmological parameters obtained by 
WMAP1+CMBsmall+2dF02 (black), WMAP3+CMBsmall plus 2dF02
(red) or plus 2dF05 (green).}
\label{fig2}
\end{figure}

\begin{figure}
\begin{tabular}{c}
\includegraphics[scale=0.4]{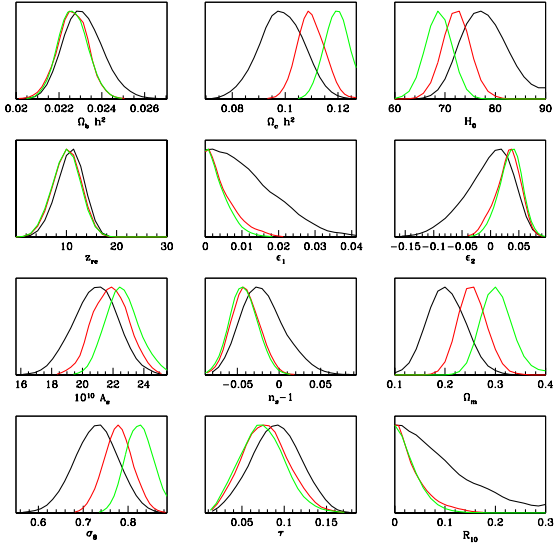}
\end{tabular}
\caption{
One dimensional marginalized
probabilities for cosmological parameters
obtained by using WMAP3+CMBsmall (black), WMAP3+CMBsmall plus 2dF02
(red) or plus SDSS (green).
}
\label{fig1}
\end{figure}

%

\begin{figure}
\begin{tabular}{cc}
\includegraphics[scale=0.41]{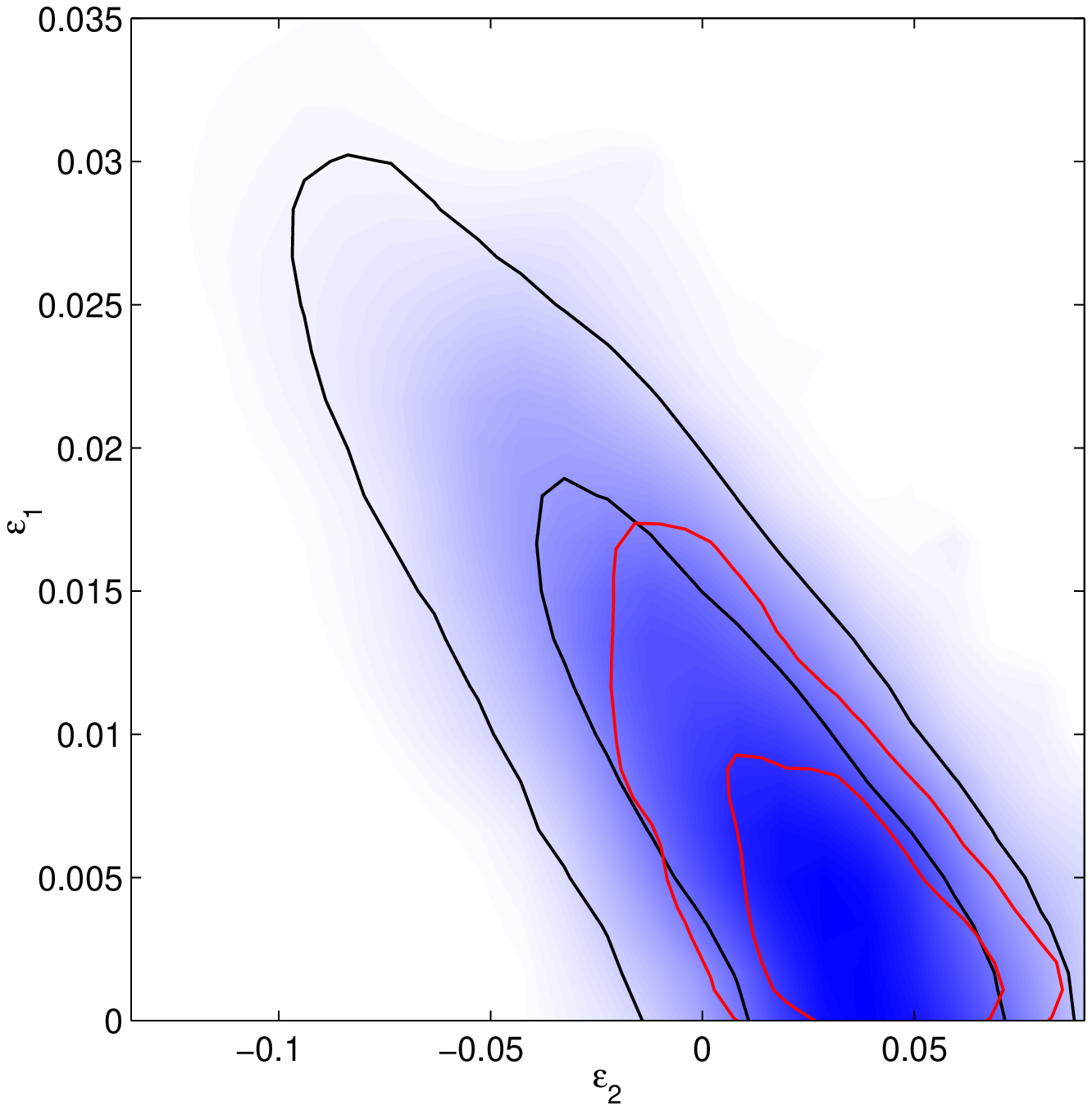}
\includegraphics[scale=0.41]{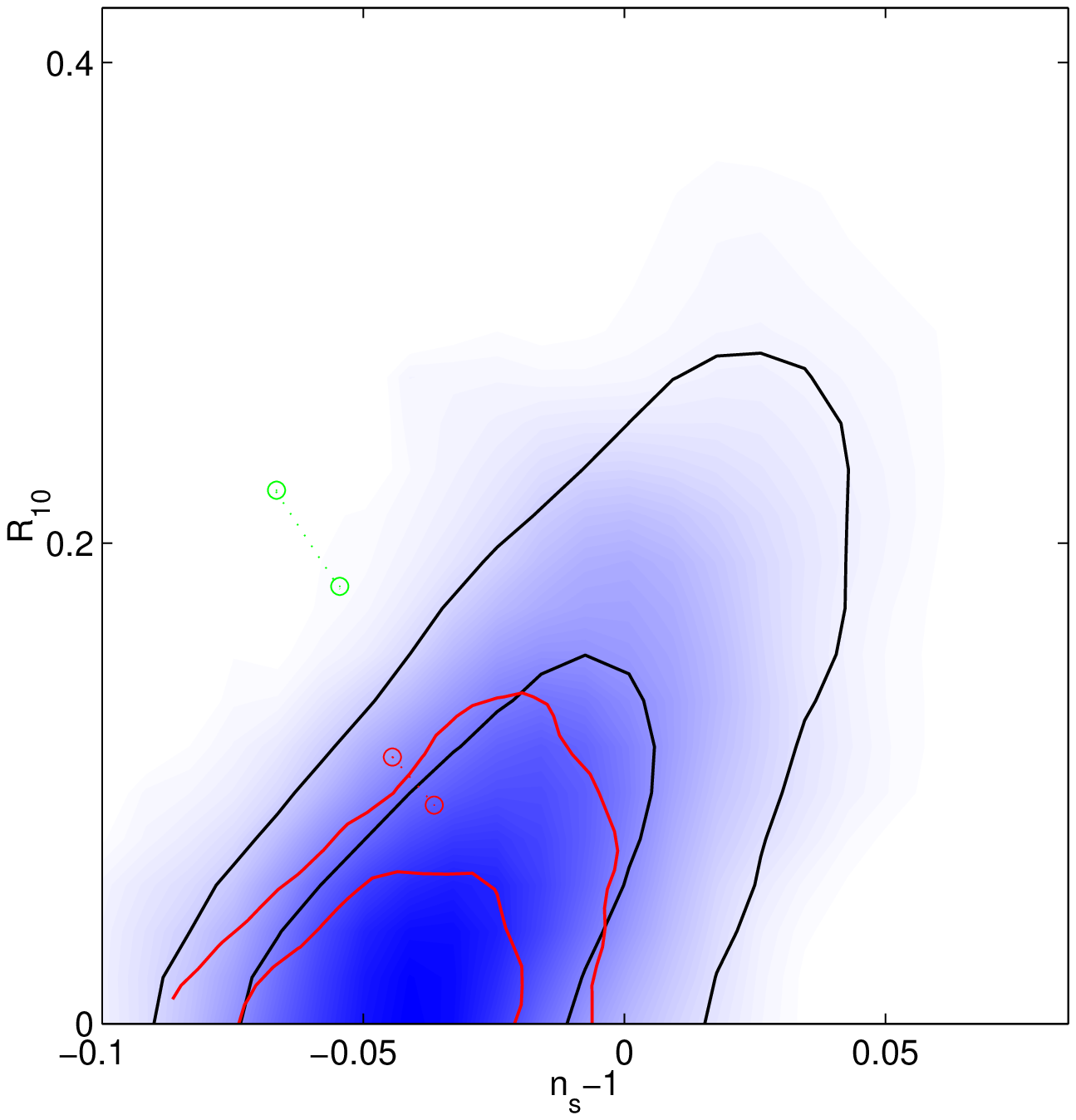}
\end{tabular}
\caption{Constraints by using WMAP3 (red) vs WMAP1 (black) plus 
CMBsmall+2dF on ($\epsilon_2 \,, \epsilon_1$) and ($n_s - 1 \,, R_{10}$)
planes at $1 \sigma$ and $2 \sigma$ level. On the right panel, 
the inflationary predictions for $V(\phi) = \lambda \phi^4/4$ (green) and 
for $V(\phi) = m^2 \phi^2/2$ (red) with $45 < \Delta N < 55$ 
are shown for comparison.}
\label{fig3}
\end{figure}

Our analysis tightens the previous bounds on $\epsilon_1$ \cite{leachliddle,MR}.
By combining WMAP3+CMBsmall with 2dF02 (SDSS) the 
$2 \sigma$ cl constraint (obtained without marginalizing on $\epsilon_2$)
is $\epsilon_1 < 0.017$ ($\epsilon_1 < 0.013$). These limits translate on the 
derived parameter $R_{10} < 0.14$ ($R_{10} < 0.10$) obtained without
marginalized on $n_S$. The current bound on $\epsilon_1$ is almost half of
the WMAP1 constraint, $\epsilon_1 < 0.031$
\footnote{Our result with WMAP1 slightly differs from the value $\epsilon_1
< 0.032$ \cite{leachliddle}, since we include also B2K in the small scale
data set, differently from \cite{leachliddle}.}.
From this result we obtain that the bound on the Hubble scale of 
inflation driven by a KG scalar field derived by
WMAP3+CMBsmall+2dF02: by using the values (not fully marginalized) 
in Fig. (\ref{fig3}) we obtain
$H/m_{\rm pl} < 1.1 \times 10^{-5}$ which improves the values previously found \cite{leachliddle,MR}
(and implies 
$(V_*^{\rm infl})^{1/4} < 2.4 \times 10^{16}$ GeV).

The impact of WMAP3 on simple monomial chaotic models with 
$V (\phi) \propto \phi^n$ strengthens the trend already 
present in WMAP1. For these models, the HFFs are given in terms of the number 
of e-folds to the end inflation $\Delta N$ as: 
$\epsilon_1 \simeq n \epsilon_2 /4 \simeq n /(4 \Delta N)$.
The quartic model is now disfavoured at more than $3 \sigma$ (although the agreement of 
the amplitude of scalar perturbations in such model is not robust with respect to the preheating mechanism \cite{FB}).
A massive inflaton is in agreement with observations, although it moves in the $2 \sigma$ cl with WMAP3 data.  

Power-law inflation \cite{pl} obtained by an exponential potential 
$V(\phi) = V_0 \exp (-\lambda \phi/M_{\rm Pl})$ 
($\lambda = \sqrt{2/p}$ with $a \sim t^p$) stay 
on the line $\epsilon_2=0$ in the ($\epsilon_2 \,, \epsilon_1$) plane. 
From the left panel of Fig. 3 and from the relation $\epsilon_1 = 1/(p-1)$ 
we obtain 
$p > 60$ at $2 \sigma$ cl (compared with $p > 53$ at $1 \sigma$ cl 
of Ref. \cite{leachliddle}). 
An interesting issue is the appearance of an upper 
bound (at $2 \sigma$ cl) starts to appear from the left panel 
of Fig. 3, i.e. $p < 500$ for WMAP3+CMBsmall+2dF02 (or SDSS): 
as is clear from Fig. 4, also this limit may depend on the 
inclusion of small scale CMB data and for its 
significativity hold the same reasoning made above for the $n_s=1$ spectrum.
Note however that the fully  
marginalized (on non-zero values of $\epsilon_2$) value for $\epsilon_1$
reported in Table I does not have any non-zero lower bound.
This upper limit for $p$, if confirmed by future data, can become important 
in constraining the number of fields which support assisted inflation 
\cite{assisted}, a particle physics realization of power-law inflation.
The constraints on the exponent of the potential at $2 \sigma$ cl are 
$0.063 < \lambda < 0.18$.  

\begin{table}
\scriptsize
\begin{tabular}{|c|c|c|c|c|}
\hline\hline
& & & \\
Parameter & WMAP1+CMBsmall+2dF02 & WMAP3+CMBsmall+2dF02 & 
WMAP3+CMBsmall+2dF05 \\
& & & \\
\hline
& & & \\
$          \Omega_b h^2 $ & $ 0.0236^{+0.0010}_{-0.0010} $    & $ 0.0226^{+0.0007}_{-0.0006} $ & 
$0.0228^{+0.0007}_{-0.0008} $\\
$       \Omega_{cdm} h^2 $ & $ 0.113^{+0.006}_{-0.006} $       & $ 0.110^{+0.005}_{-0.005} $ & 
$0.105^{+0.005}_{-0.005}$ \\
$            H_0        $ & $ 73.4^{+1.3}_{-1.7} $            & $ 72.3^{+1.1}_{-1.2} $ & 
$ 74.3^{+2.5}_{-2.3} $ \\
$            z_{re}     $ & $ 11.7^{+1.9}_{-1.5} $            & 
$ 10.1^{+1.2}_{-1.1} $ & 
$ 10.0^{+1.6}_{-0.9} $ \\
$           \epsilon_1  $ & $ < 0.024 $    & $ < 0.013 $ & 
$ < 0.016 $ \\
$           \epsilon_2  $ & $ 0.001^{+0.038}_{-0.037} $       & 
$ 0.03^{+0.02}_{-0.02} $ & 
$ 0.02^{+0.02}_{-0.02} $ \\
$    10^{10}A_{\rm s}   $ & $  22.5^{+ 1.6}_{- 1.6} $         & $ 21.8^{+1.2}_{-1.1} $ & 
$ 21.3^{+1.3}_{-1.3} $ \\
$          n_{\rm s}-1  $ & $ -0.021^{+0.025}_{-0.026} $      & 
$ -0.040^{+0.017}_{-0.017} $ & 
$ -0.035^{+0.007}_{-0.009} $ \\
$          \Omega_m     $ & $ 0.256^{+0.032}_{-0.031} $       & $ 0.255^{+0.025}_{-0.025} $ & 
$ 0.232^{+0.020}_{-0.021} $ \\
$          \sigma_8     $ & $ 0.82^{+0.04}_{-0.04} $          & $ 0.78^{+0.03}_{-0.03} $ & 
$ 0.76^{+0.03}_{-0.04} $ \\
$              \tau     $ & $ 0.10^{+ 0.04}_{- 0.04} $        & $ 0.078^{+0.027}_{-0.026} $ & 
$ 0.080^{+0.031}_{-0.034} $ \\
$            R_{10}     $ & $ < 0.22 $      & $ < 0.10 $ &
$ < 0.13 $ \\
\hline\hline
\end{tabular}
\caption{Mean values and $1 \sigma$ constraints for the $7$ basic 
parameters and other $5$ 
derived quantities from the first and 
three year WMAP release combined with the two release of 2dF plus CMBsmall. For $\epsilon_1$ and 
$R_{10}$ the 2$\sigma$ upper bounds are given.
\label{tab:base} }
\end{table}

\begin{figure}
\begin{tabular}{cc}
\includegraphics[scale=0.41]{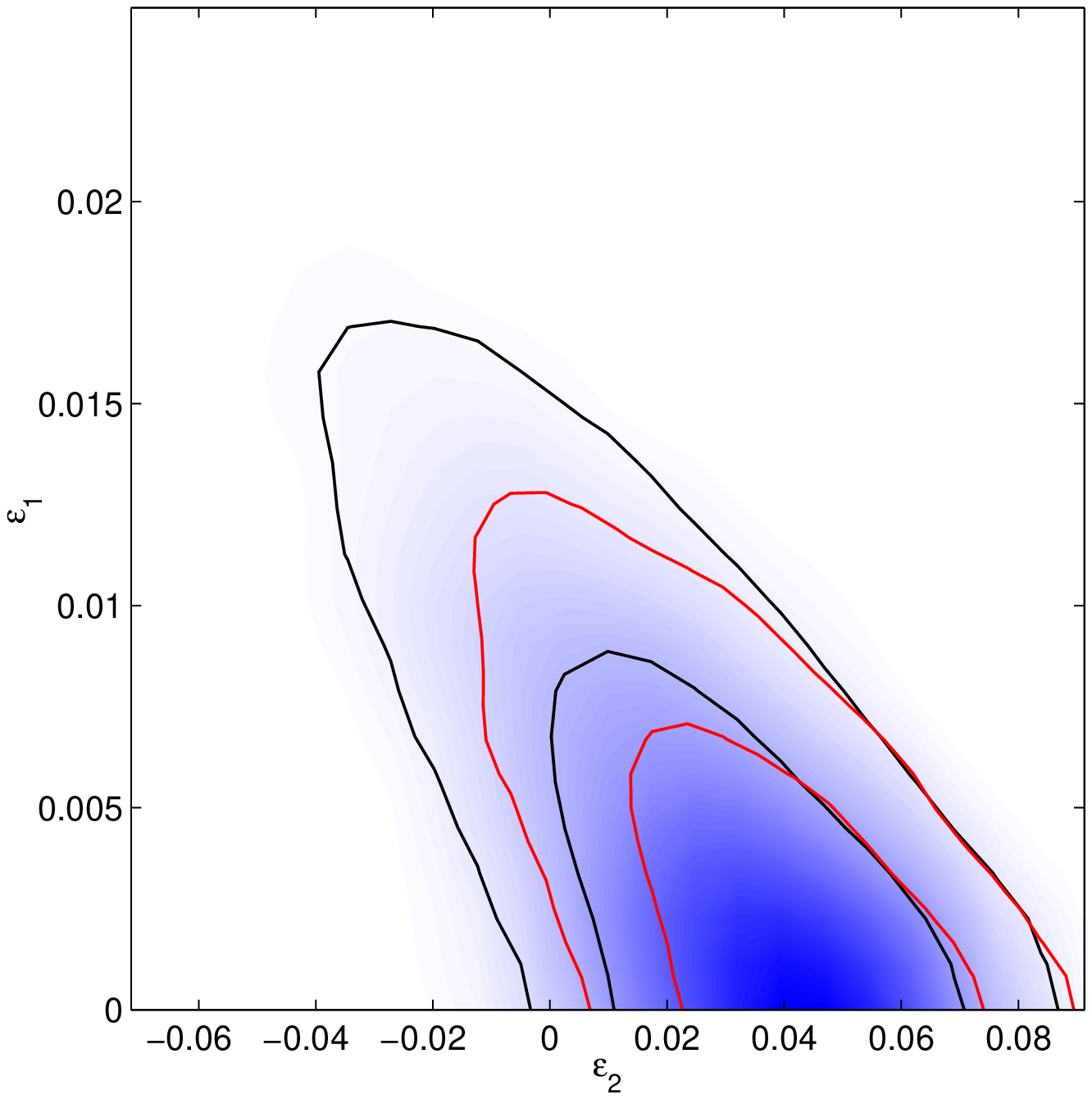}
\includegraphics[scale=0.41]{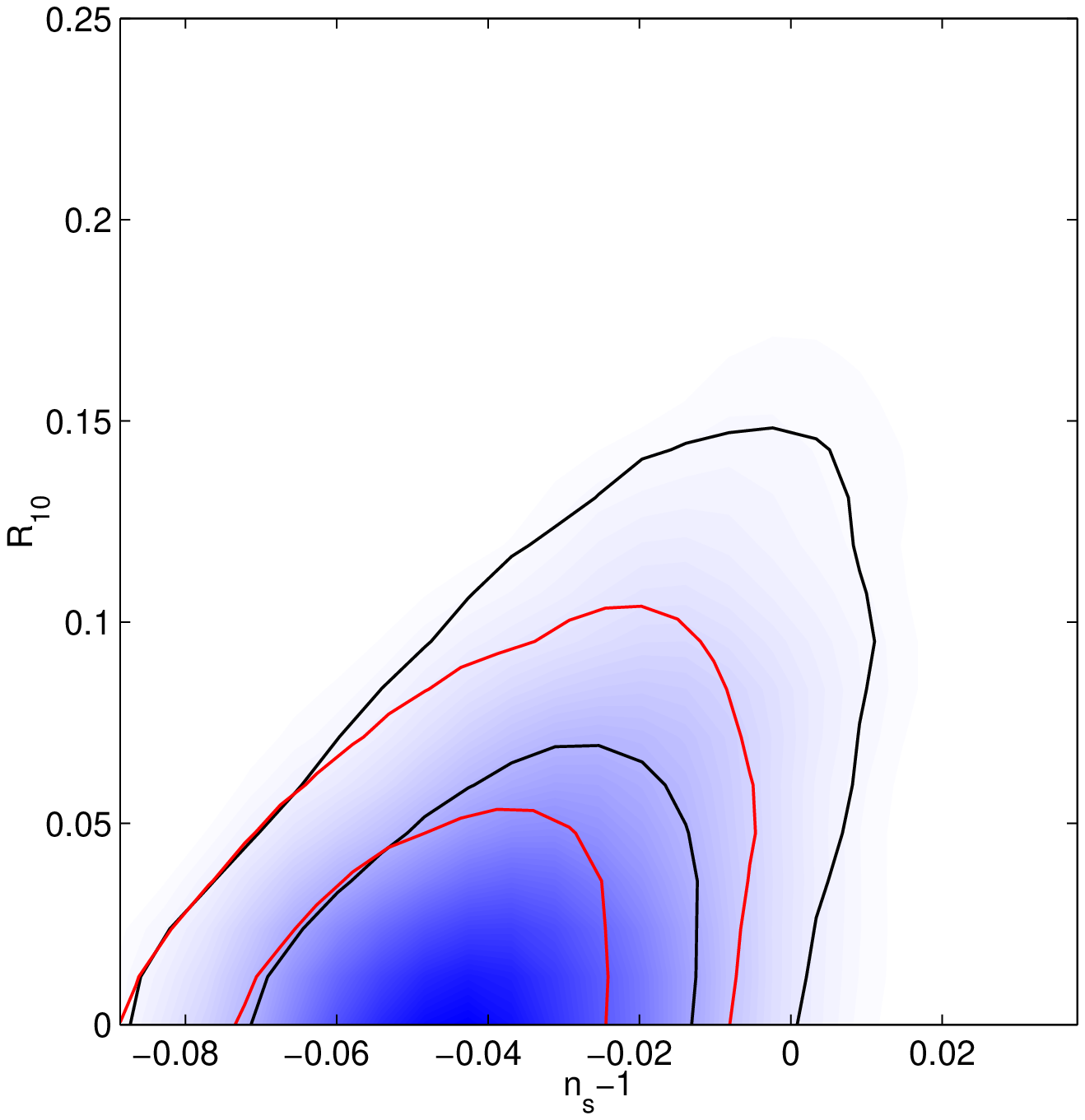}
\end{tabular}
\caption{Constraints by using WMAP3+SDSS (black) vs WMAP3+CMBsmall+SDSS 
(red) on ($\epsilon_2 \,, \epsilon_1$) and ($n_s - 1 \,, 
R_{10}$) planes at $1 \sigma$ and $2 \sigma$ level. The right panel  
shows how the inclusion of small scale CMB data drives 
the spectral scalar index towards red values, disfavouring at just 
2$\sigma$ cl the $n_s=1$ line. The constraints on $n_S$ 
we obtain from WMAP3+SDSS are 
in complete agreemente with \cite{KKMR,PE}.} \label{fig3bis}
\end{figure}

\section{Relaxing the standard consistency condition}

We now discuss the bounds obtained by relaxing the consistency condition 
(\ref{ttos}) and by therefore varying $8$ parameters (the $7$ of Eq. 
(\ref{string}) plus the tensor-to-scalar ratio $r_{0.01}$ at the pivot 
scale $k_* = 0.01 {\rm Mpc}^{-1}$). 
From a theoretical point of view, this means that we are not specifying the 
Lagrangian $p (\phi, X)$ for the inflaton $\phi$ and its kinetic term 
$X = - \nabla_\mu \phi \nabla^\mu \phi / 2$, but we are 
restricting ourselves to inflationary spectra at 
first order in HFF. In this way the consistency condition in Eq. (\ref{ttos}) 
is modified by the speed of sound $c_S$ for the inflaton \cite{GM}:
\be
\frac{P_{0 \, T}}{P_{0 \, S}} = - 8 c_S n_T \,, \quad {\rm with} \quad c_S^2 = \frac{\partial 
p}{\partial X} \,,
\label{speed}
\ee
whereas the spectral indices in Eqs. (\ref{eqn:bs2},\ref{eqn:bt2}) are not modified by $c_S$ at first 
order in HFF for constant $c_S$ \cite{WCW}. Our analysis with a tensor-to-scalar ratio unrelated to the tensor 
spectral index also applies to inflation driven by more than one scalar field, by assuming only curvature 
perturbations after nucleosynthesis (i.e. that isocurvature perturbations generated during inflation have 
completely transferred to the adiabatic mode when 
fluctuations are initialized in the Einstein-Boltzmann code).

\begin{figure}
\begin{tabular}{cc}
\includegraphics[scale=0.3]{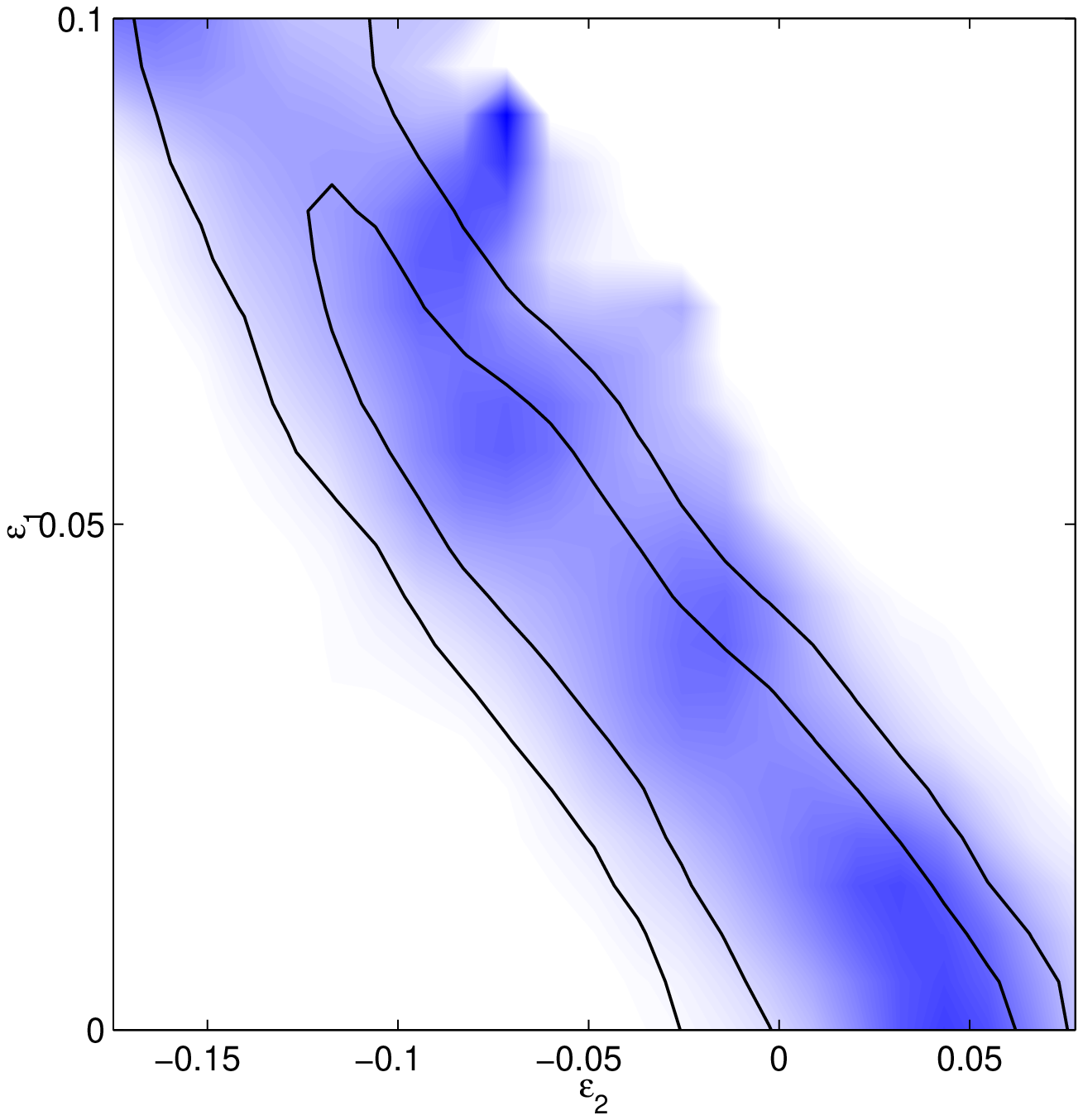}
\includegraphics[scale=0.3]{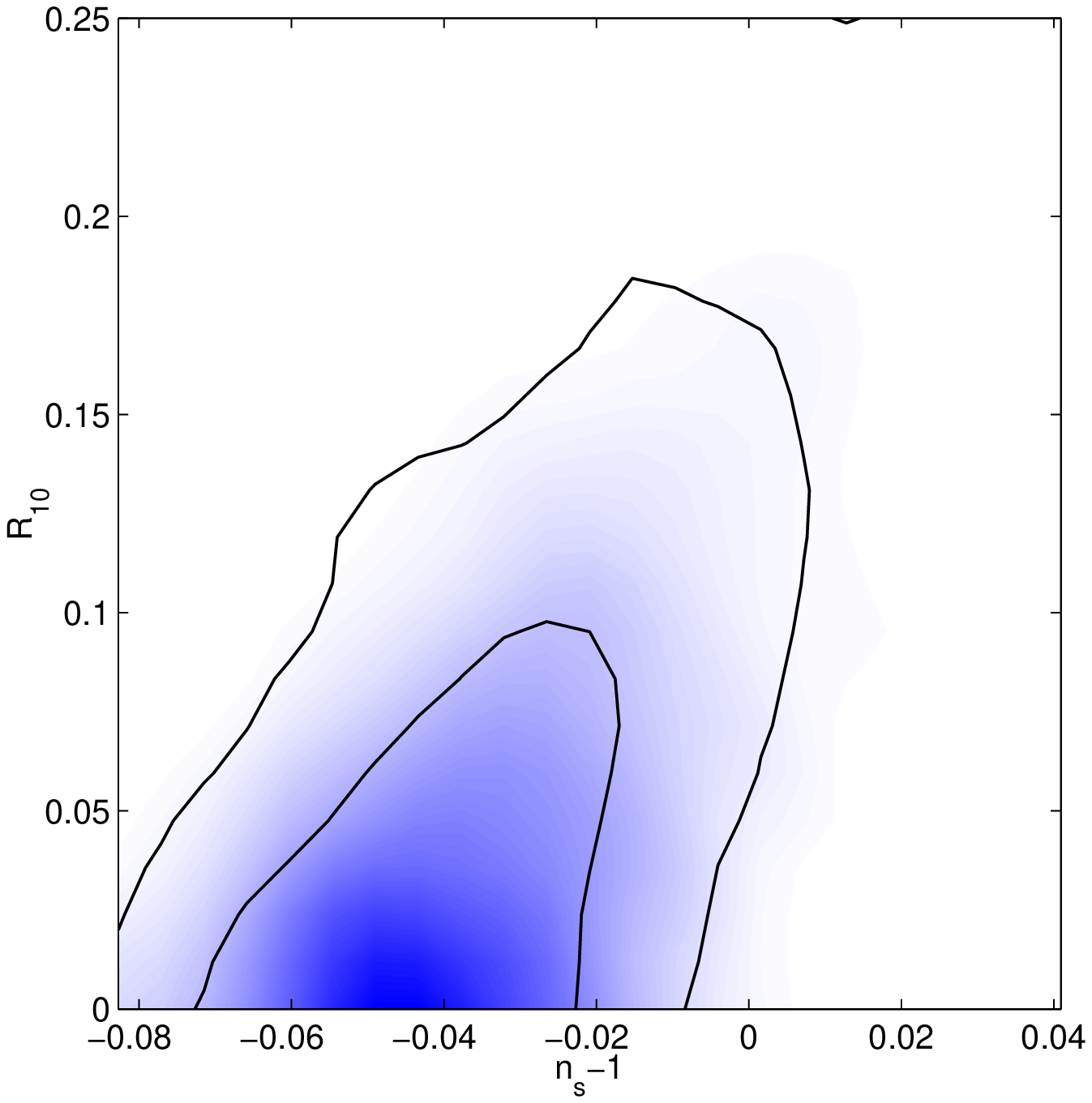}
\end{tabular}
\begin{tabular}{cc}
\includegraphics[scale=0.3]{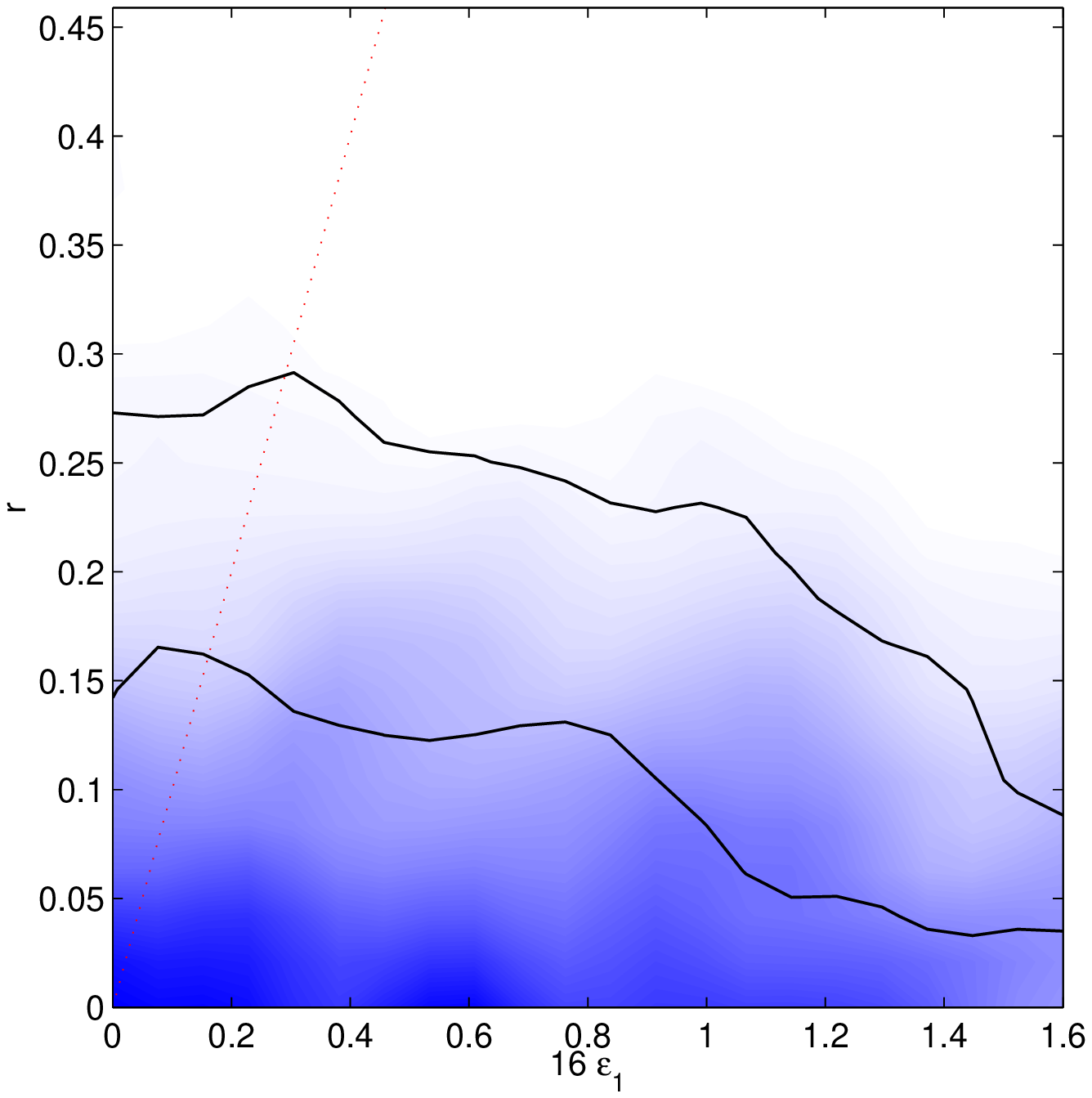}
\end{tabular}
\caption{Two dimensional contours at $1 \sigma$ and $2 \sigma$ level 
obtained by WMAP3+CMBsmall+2dF02. 
The plots reported from left to right are $(\epsilon_2\, \epsilon_1)$, 
$(n_S - 1 \,, R_{10})$, $(16 \, \epsilon_1 \,, r)$, respectively.
In the plane $(16 \, \epsilon_1 \,, r)$, the red line describing the 
consistency condition $r=16 \epsilon_1$ is also drawn for comparison.}
\label{fig4}
\end{figure}

The two dimensional constraints for relevant parameters are presented in 
Fig. (\ref{fig4}). The constraints on the ($\epsilon_1 \,, \epsilon_2$) 
plane become looser than in the previous section, since $\epsilon_1$ is 
now unrelated to the tensor-to-scalar ratio.
Without marginalizing on the scalar spectral index, we obtain 
$r < 0.29$ (or $R_{10} < 0.185$) at 2$\sigma$ cl, 
a bound weaker than the one obtained by imposing the standard 
consistency condition in Eq. (\ref{ttos}) ($r < 0.27$ or $R_{10} < 0.14$, 
see previous section). Note that inflation driven by KG scalar field 
(i.e. $c_S=1$) is fully consistent with present data.

\section{NEC violating Inflation} 

Inflation may occur with a violation of the null energy condition (NEC) 
\cite{BFM}, i.e. with $\epsilon_1 < 0$. Since $n_T = -2 \epsilon_1$, 
a prediction of NEC violating inflation is a blue spectrum for gravitational 
waves. A similar blue spectrum for gravitational waves may occur 
also in scalar-tensor theories and $f(R)$ gravity theories 
which are NEC violating in the Einstein frame conformally related 
to the original Jordan one.
In order to remain with $3$ inflationary parameters, we shall 
restrict ourselves to a simple model, just changing the sign of the 
kinetic term of standard inflation, 
which leads to relation $r \simeq - 16 \epsilon_1$ in the analysis \cite{piao}. 
The prior used is $-10^{-4} < \epsilon_1 < -0.1$. 

\begin{figure}
\begin{tabular}{c}
\includegraphics[scale=0.4]{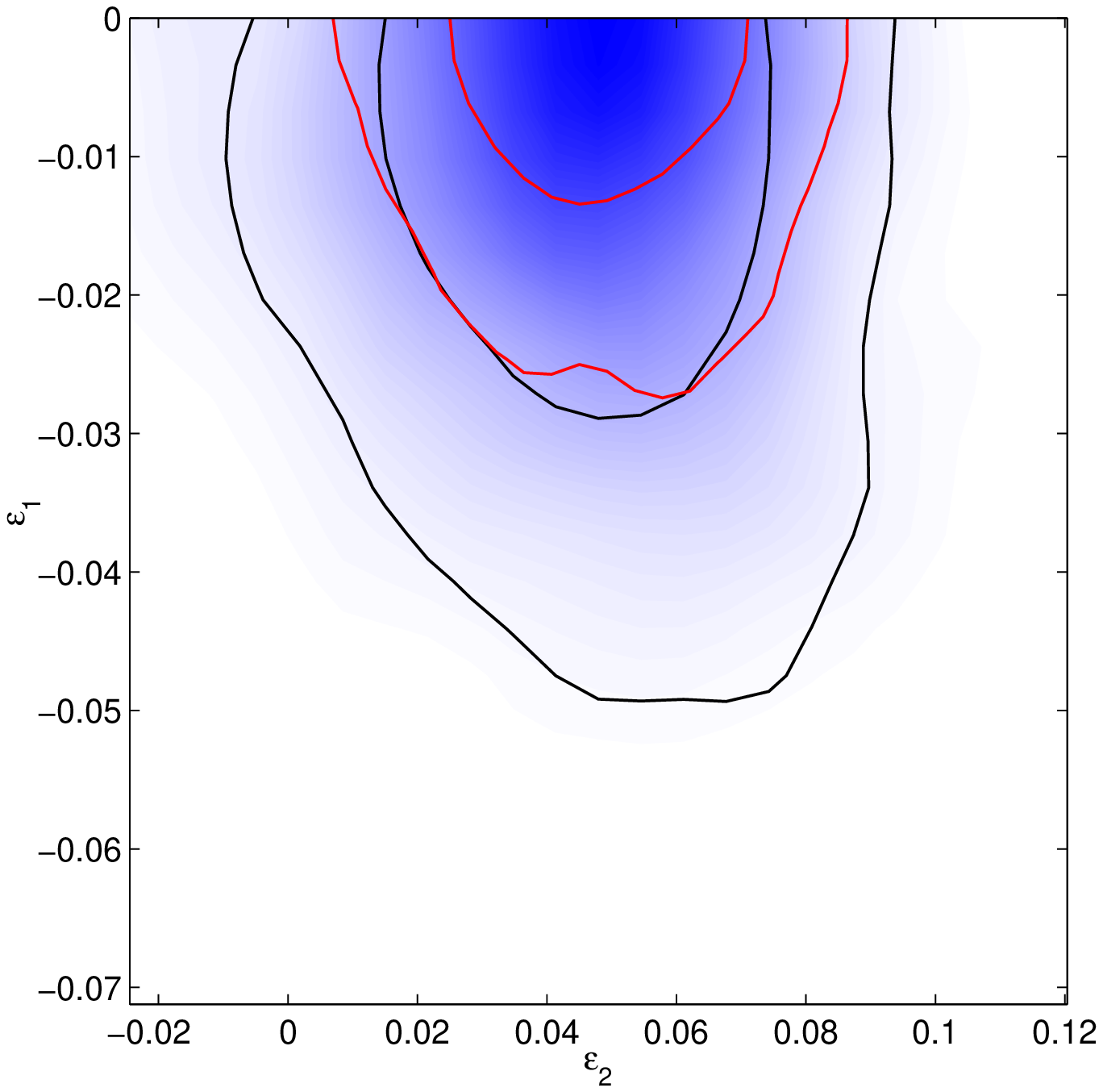}
\includegraphics[scale=0.4]{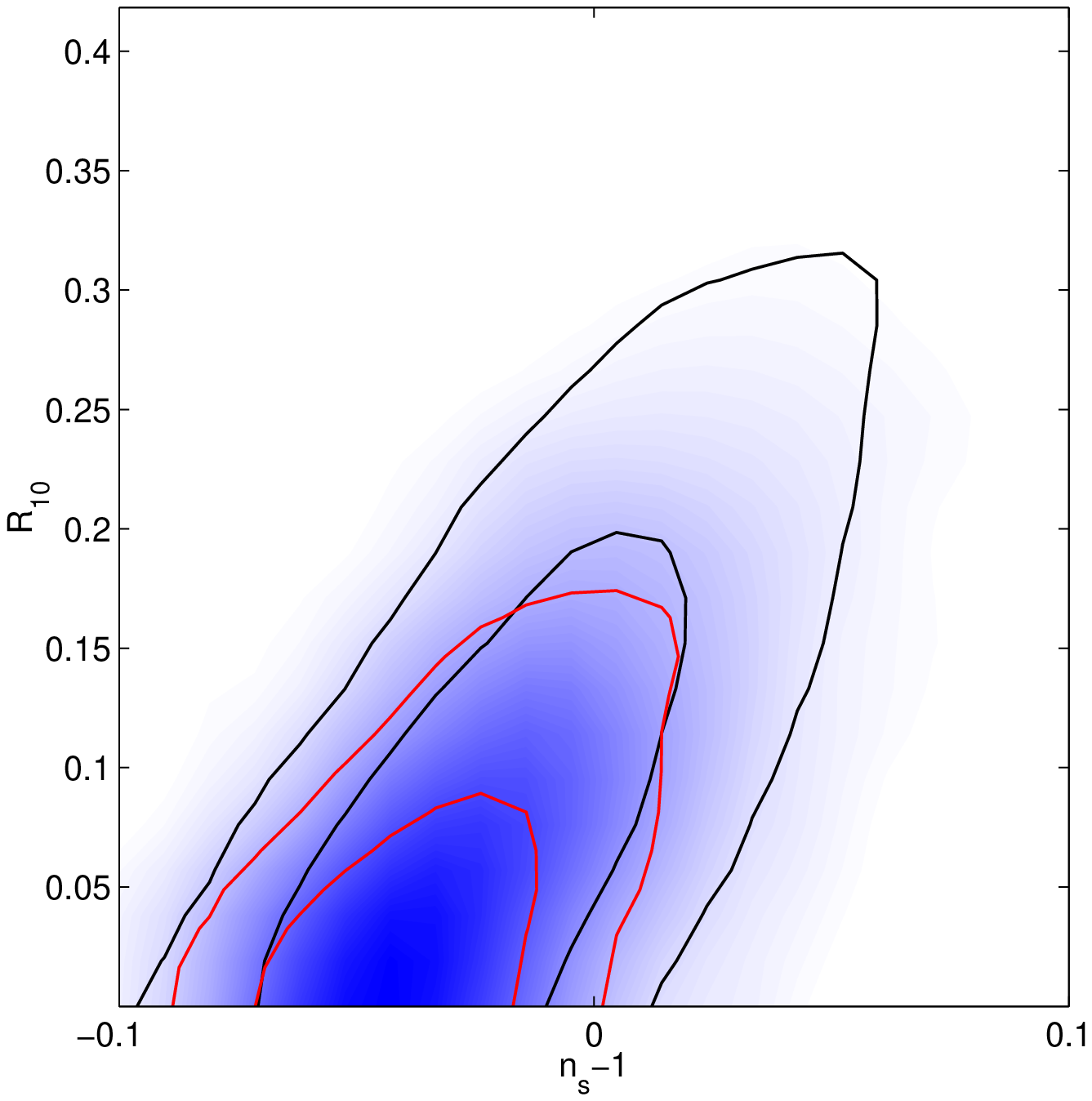}
\end{tabular}
\caption{WMAP3 (red) vs WMAP1 (black) $1 \sigma$ and $2 \sigma$ contours 
on ($\epsilon_2 \,, \epsilon_1$) (left) and ($R_{10} \,, n_s-1$) for NEC 
violating inflation. The other data used are CMBsmall and 2dF02.}
\label{fig5}
\end{figure}

\begin{figure}
\begin{tabular}{c}
\includegraphics[scale=0.5]{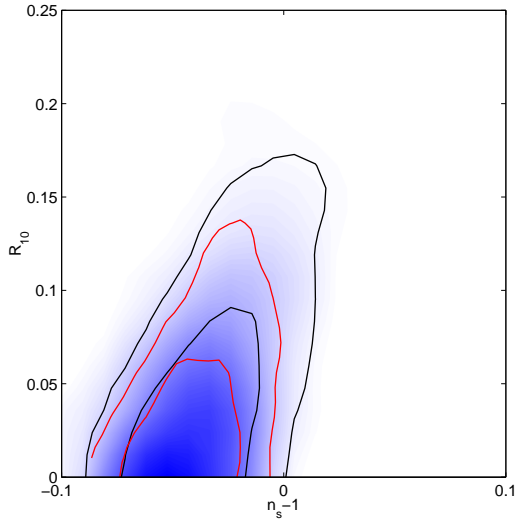}
\end{tabular}
\caption{Standard inflation (red) vs NEC violating inflation (black) on 
the ($n_S - 1 \,, R_{10}$) at $1 \sigma$ and $2 \sigma$ level. }
\label{fig6}
\end{figure}

The constraints on the ($\epsilon_2 \,,\epsilon_1$) for 
$\epsilon_1 < 0$ are presented in the left panel of Fig. (\ref{fig5}). 
The contours on ($\epsilon_2 \,,\epsilon_1$) for NEC violated 
inflation are completely different from standard inflation, although 
the amount of tensor is slightly loosely constrained with respect to 
standard inflation because of the the 
blue slope for gravitational waves, as can be seen in Fig. (\ref{fig6}).
The best fits for standard inflation and for NEC violating inflation have the same 
value for $\chi^2$ ($=5673$) with WMAP3+CMBsmall+2dF02.

From the ($\epsilon_2 \,,\epsilon_1$) plane, it can be seen that an 
exponential potential (described by the $\epsilon_2=0$ line) is disfavoured  
at more than $2 \sigma$ cl. 
A potential 
$V (\phi) \sim \exp \left( \bar{\epsilon}_2 \phi^2/M_{\rm pl}^2 \right)$ 
with $0.02 \lesssim \bar{\epsilon}_2 \lesssim 0.08$ is 
consistent with the data.

\section{Second Order HFF Results}
 
We now allow $b_{{\rm S} \, 2} \,, b_{{\rm T} \, 2} \ne 0$ 
in Eq. (\ref{plex}), which, according to inflationary 
predictions, is equivalent to take into account terms which 
are quadratic in HFFs. Inflationary models in which the slow-roll 
condition is not well satisfied can be described analitycally with better accuracy by 
including the spectra at second order in HFFs \cite{LLMS,CFKLV,makarov}. 

Let us first try to predict what may happen by inserting the PSs in 
Eq. (\ref{plex}) which allow running with respect to 
the pure power-law PSs analyzed in Section III: at the price of 
one additional parameter, the input scalar and tensor PSs are now 
paraboles in $\ln (k/k_*)$, and not straight lines. It is expected that the $7$ 
parameters used in Section III change their mean value and broaden 
their variance in presence of the running, the additional parameter.
Having in mind the  
degeneracy in cosmological parameters, it is also possible  
that the best-fit model with runnings may be fairly different 
from the best-fit model without runnings. 
Within our analysis, we shall see that both these possibilities indeed 
occur with different LSS data sets.

WMAP 3yr analysis included the running of the scalar spectral index as an 
additional parameter \cite{spergel}, 
neither enforcing a self-consistent tensor-to-scalar ratio to second order 
in HFFs nor including a running for the tensor 
spectral index. See also other studies of the 
running of the spectral index for WMAP1 \cite{clinehoi} 
and WMAP3 \cite{EP}.

\begin{figure}
\begin{tabular}{c}
\includegraphics[scale=0.14]{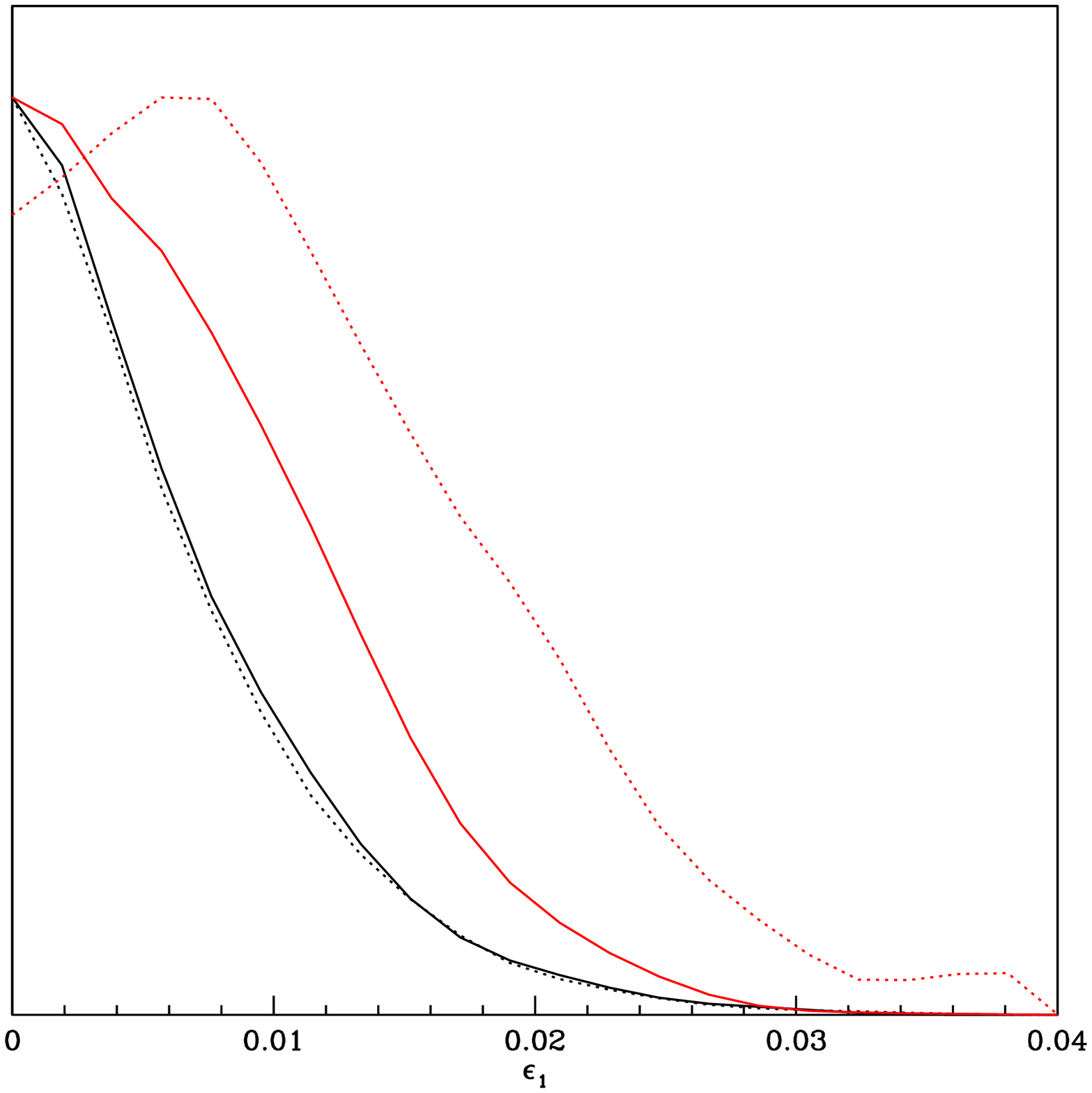}
\includegraphics[scale=0.14]{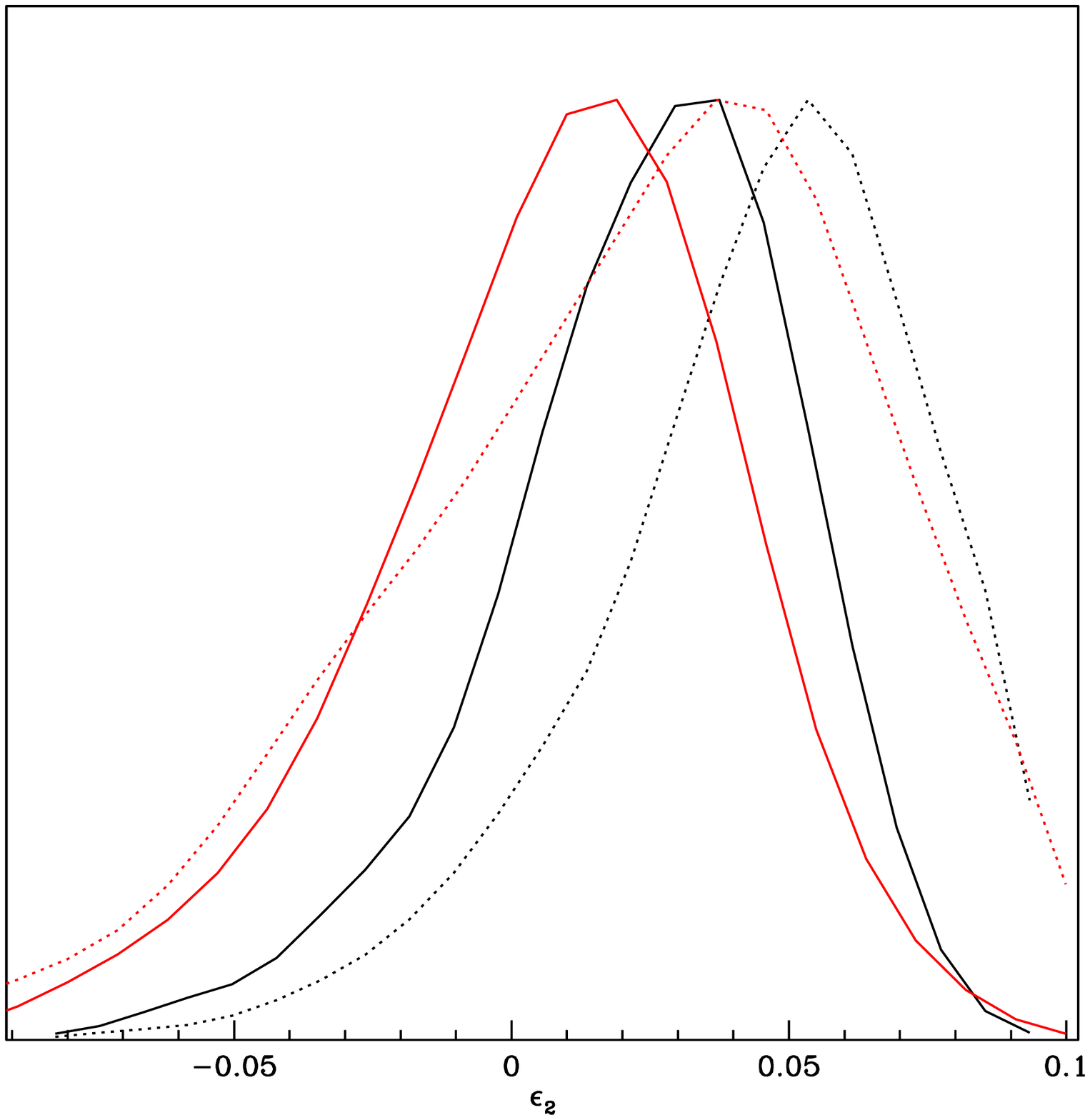}
\includegraphics[scale=0.14]{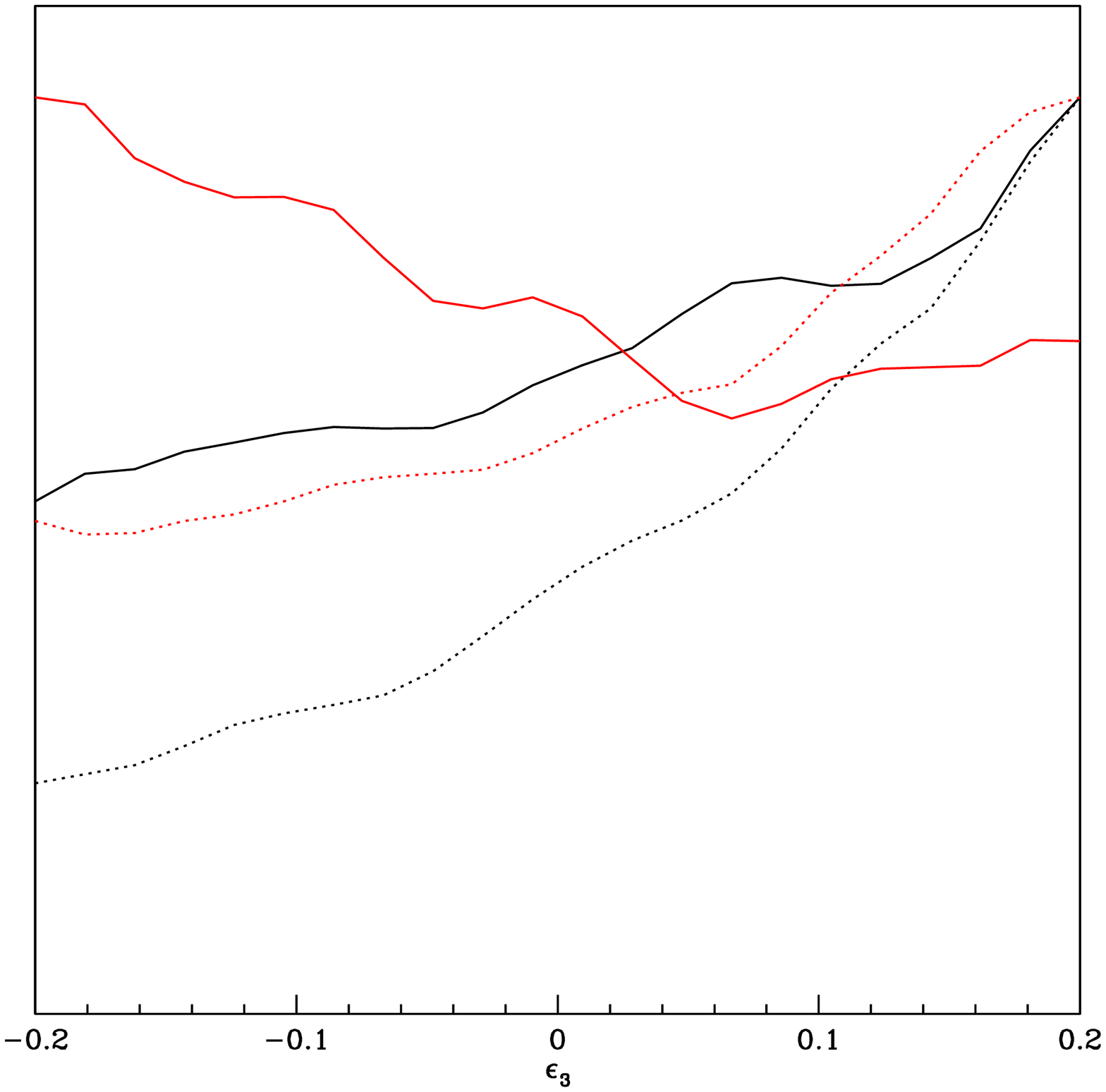}
\end{tabular}
\caption{One dimensional mean likelihhods (dashed) and marginalized 
probabilities (solid)
for $\epsilon_1 \,, \epsilon_2 \,, \epsilon_3$ using a prior 
$-0.2 \le \epsilon_3 \le 0.2$ and the GFM. 
We show for comparison results from WMAP1 
(red) and WMAP3 (black) plus CMBsmall+2dF02.}
\label{fig7}
\end{figure}

We show in Fig. (\ref{fig7}) the mean likelihoods and marginalized 
probabilities for $\epsilon_1 \,, \epsilon_2 \,, \epsilon_3$ with 
a prior $[-0.2 \,, 0.2]$ on $\epsilon_3$ 
obtained by the two WMAP releases. The results obtained are 
qualitatively in agreement with \cite{leachliddle} (for WMAP1) and 
\cite{MR} (for WMAP3), although the sets of auxiliary data used are not 
the same. It is clear from Fig. (\ref{fig7}) 
and Eqs. (\ref{eqn:bs2}) that such prior constrains $|\alpha_S| 
\lesssim 0.01$, although there is a tendency to favour positive values 
for $\epsilon_3$, in particular for WMAP3 (as also pointed out in 
\cite{MR}).  

\begin{figure}
\begin{tabular}{c}
\includegraphics[scale=0.14]{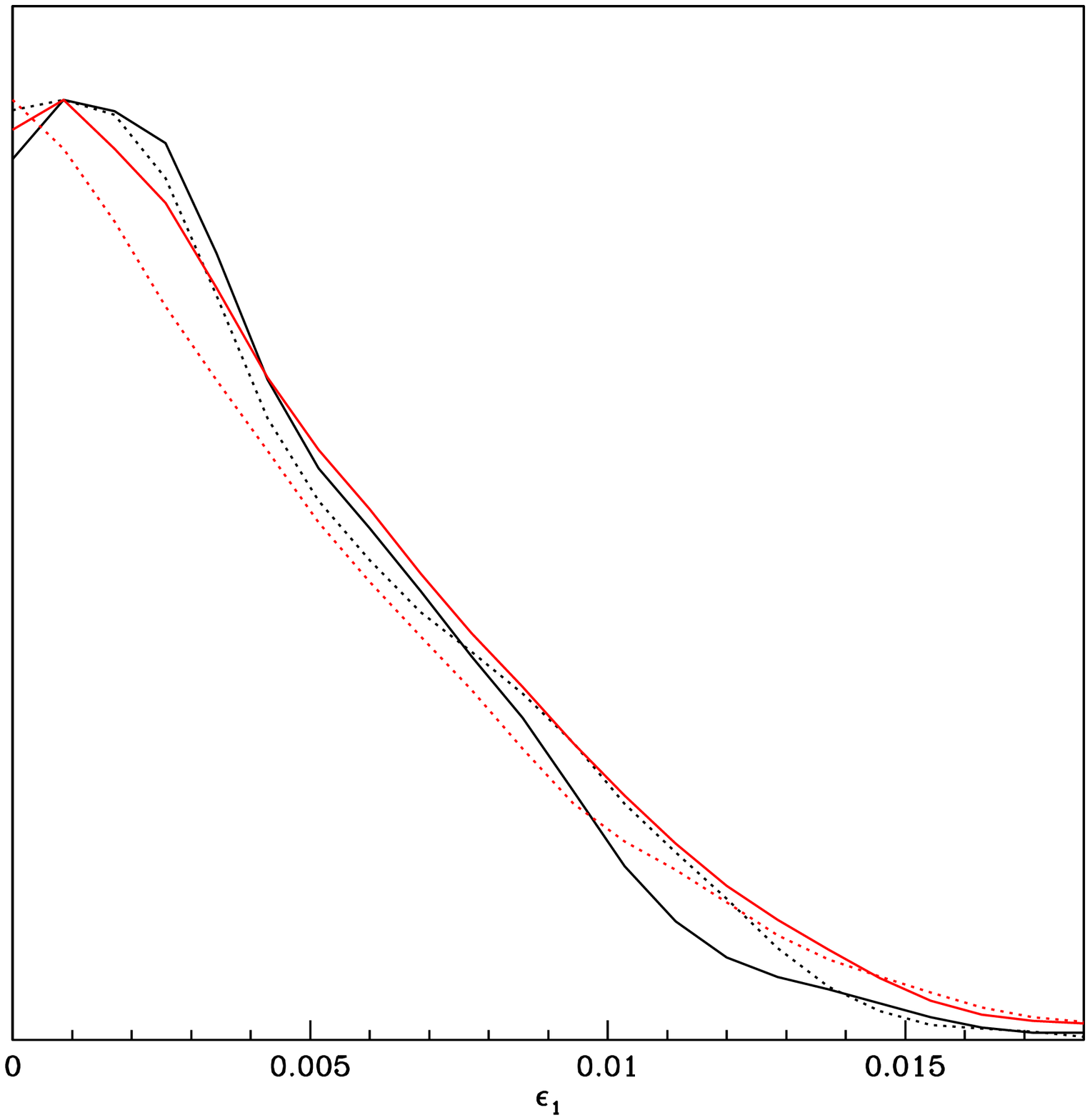}
\includegraphics[scale=0.14]{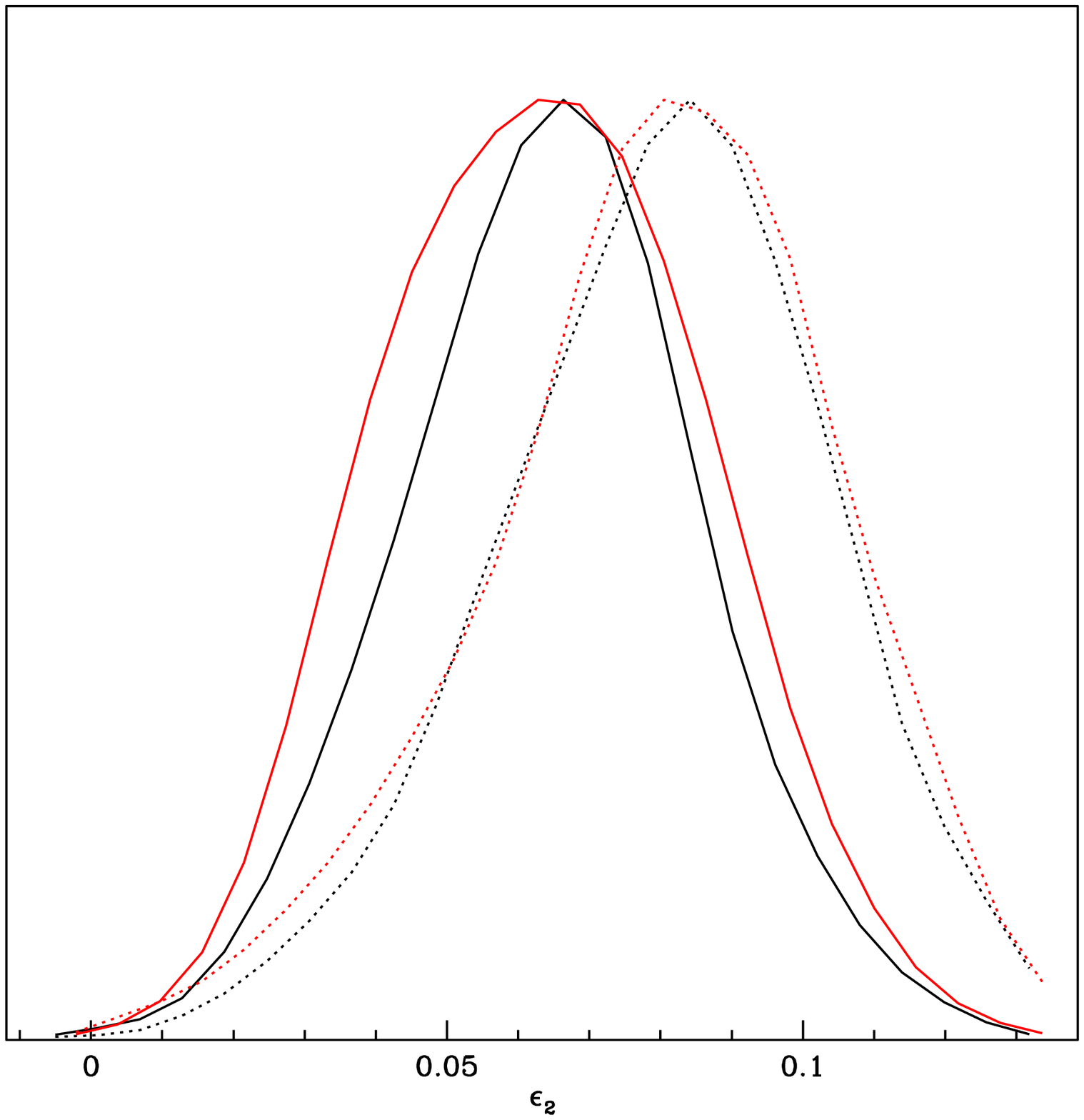}
\includegraphics[scale=0.14]{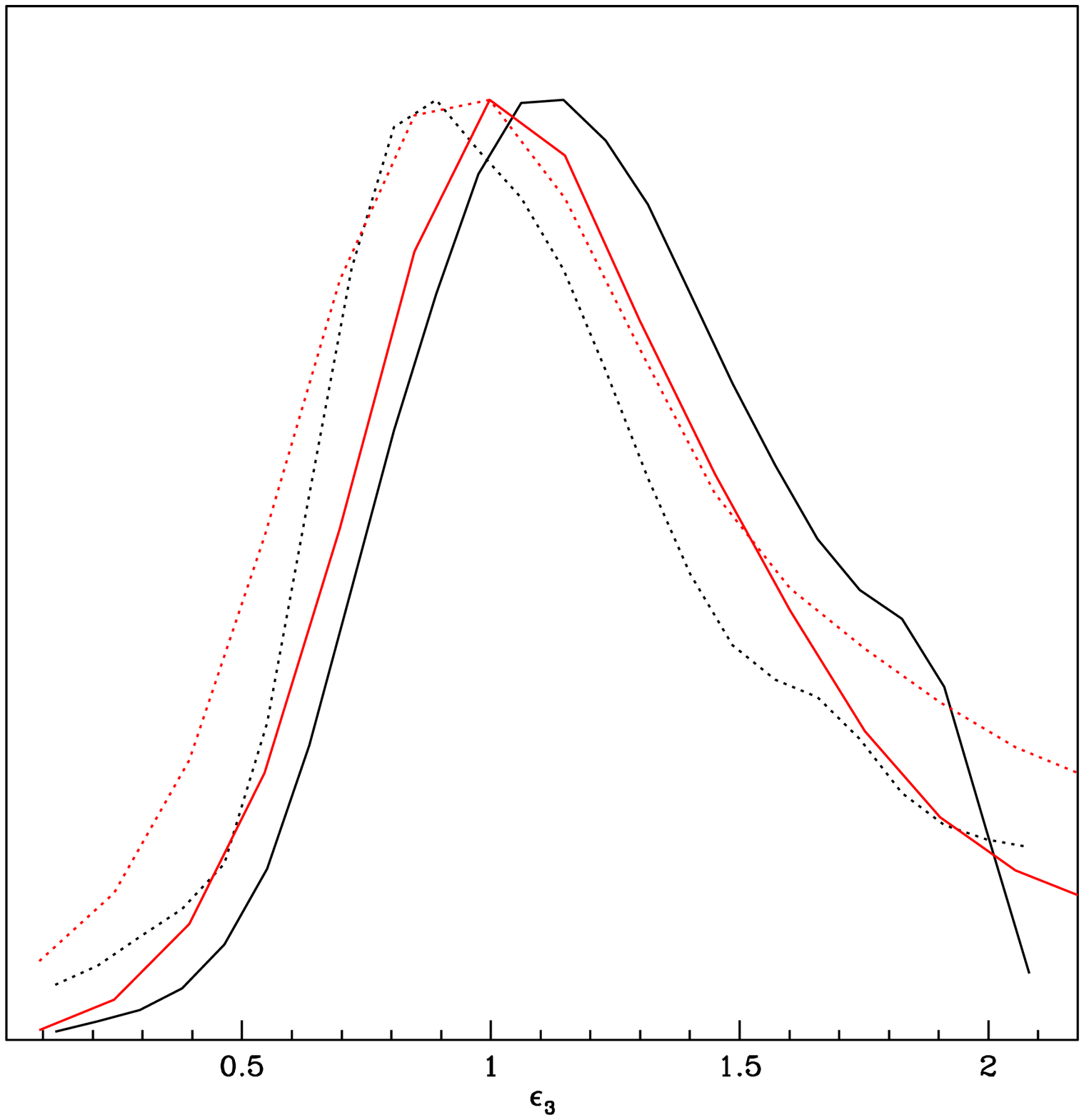}
\includegraphics[scale=0.20]{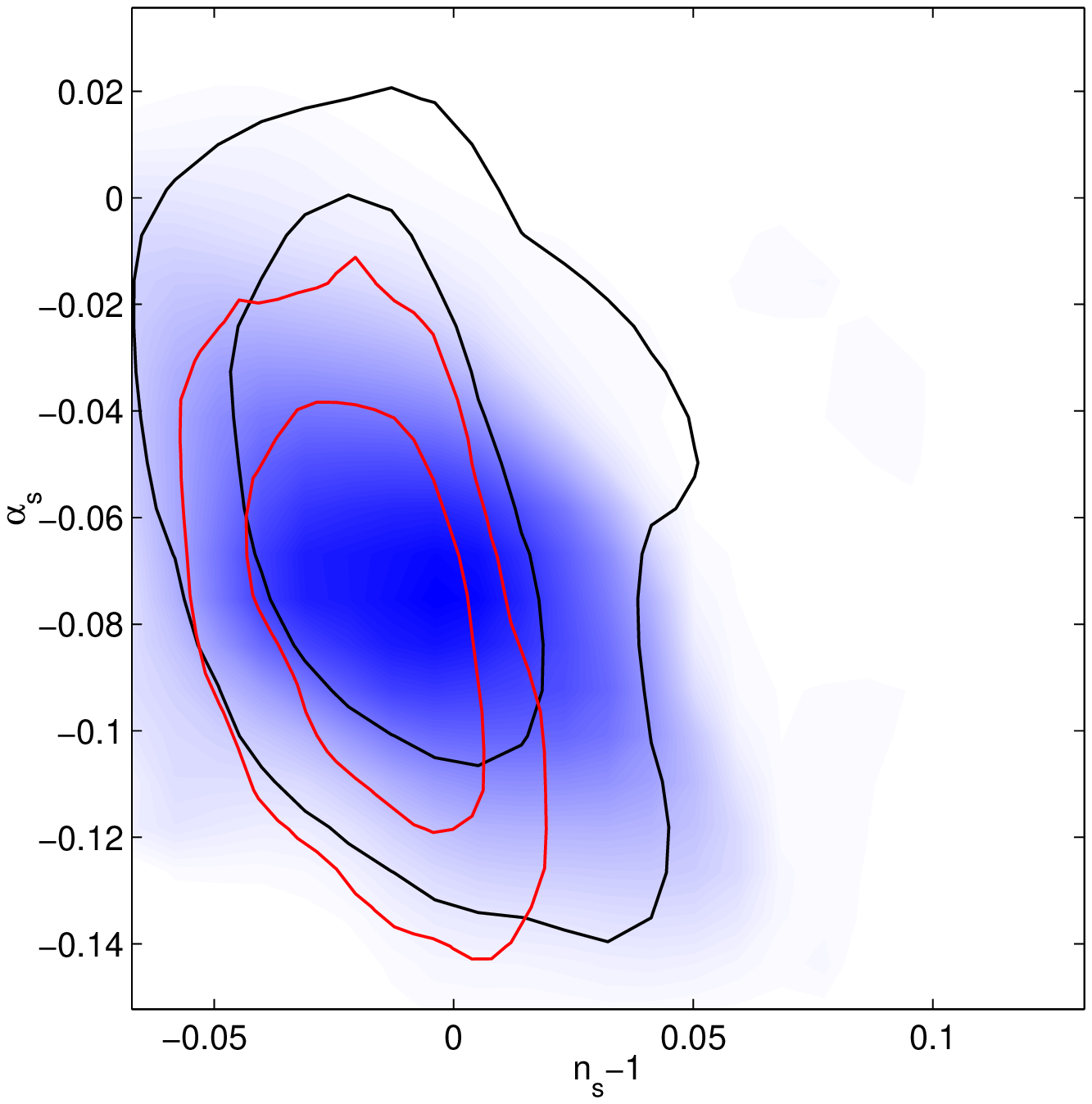}
\end{tabular}
\caption{First three panels: 
one dimensional mean (dashed) and marginalized (solid)
likelihoods for $\epsilon_1 \,, \epsilon_2 \,, \epsilon_3$ (left, middle 
and right panel, respectively) obtained by using the two different 
analytic approximation for the inflationary power spectra with 
WMAP3 + CMBsmall + SDSS: GFM (black) and MCE (red). The last panel on 
the right shows the two dimensional constraints on the 
($n_S - 1, \alpha_S$) plane by using the GFM for WMAP1 (black) and 
WMAP3 (red) plus CMBsmall and SDSS.}
\label{fig8}
\end{figure}

In order to have a non trivial result for $\epsilon_3$ it is crucial to
consider a broader prior on this parameter. The justification for 
this broader prior for $\epsilon_3$ is its appearance in 
Eqs. (3-10) not at 
linear level, but always multiplied by $\epsilon_2$.
The second order formalism is limited by 
$\epsilon_1 \epsilon_2 \,, \epsilon_2 \epsilon_3 << 1$ and not by 
$\epsilon_3 << 1$ (see also \cite{makarov} for a similar claim). 
By repeating the same analysis with a much broader prior \footnote{We 
have used $[-1,2]$ for WMAP3+SDSS and $[-2,5]$ for WMAP1+SDSS and 
WMAP3+2dF.}, we find more definite preferred values around 
$\epsilon_3 \sim 1$, as can be seen from Fig. (\ref{fig8}). 
In the right panel of 
Fig. (\ref{fig8}) we show how the statistical
evidence for running has increased from WMAP1 to WMAP3, although
remaining still weak.

The dependence of $\ln (P(k)/P_0 (k_*))$ in Eq. (\ref{plex}) 
and of the coefficients $b_{{\rm S \,, T} i}$ in Eqs. (4-10) on 
$d \epsilon_2 / d N$ ($= \epsilon_2 \epsilon_3$) is the reason 
for a very inefficient exploration of runnings of the spectral indexes by 
the MCMC with basic parameter $\epsilon_3$. 
We find much more efficient sampling directly on $d 
\epsilon_2/ d N = \epsilon_2 \epsilon_3$ rather than on $\epsilon_3$. We 
present results on chains obtained sampling on $\epsilon_2 \epsilon_3$ 
with prior $[-0.5 \,, 0.5]$ in Table 2 and Figs. 9-10.

\begin{table}
\scriptsize
\begin{tabular}{|c|c|}
\hline
 & \\
Parameter 
& WMAP3+CMBsmall+2dF05 \\
& \\
\hline
& \\
$          \Omega_b h^2 $ & 
$ 0.0218^{+0.0009}_{-0.0008} $
\\
$       \Omega_{cdm} h^2 $ 
& $0.112^{+0.006}_{-0.006}$ \\
$            H_0        $ & 
$ 71.0^{+2.7}_{-2.7} $ \\
$            z_{re}     $ & 
$ 12.6^{+1.7}_{-1.0} $ \\
$           \epsilon_1  $ 
&  $ < 0.031 $ \\
$           \epsilon_2  $ & 
$ 0.015^{+0.020}_{-0.018} $ \\
$    \epsilon_2 \epsilon_3 $ & 
$ 0.076^{+0.033}_{-0.033} $ \\
$    10^{10}A_{\rm s}   $ & 
$ 23.0^{+1.7}_{-1.7} $ \\
$          n_{\rm s}-1  $ & 
$ 0.010^{+0.038}_{-0.036} $ \\
$          \alpha_{\rm s}  $ & 
$ - 0.076^{+0.032}_{-0.033} $ \\
$          \sigma_8     $ & 
$ 0.77^{+0.04}_{-0.03} $ \\
$              \tau     $ & 
$ 0.10^{+0.04}_{-0.03} $ \\
$            R_{10}     $ & 
$ < 0.40 $ \\
\hline
\end{tabular}
\caption{Mean values and $1 \sigma$ constraints for the $8$ basic
parameters and other $5$
derived quantities from 
WMAP three year data + CMBsmall + 2dF05. Note that 
for $\epsilon_1$ and $R_{10}$ 
the 2$\sigma$ upper bounds are given. The $2 \sigma$ constraints on 
running of the spectral index is: $-0.141 < \alpha_{\rm S} < -0.013$.
\label{tab:base2} }
\end{table}

\begin{figure}
\begin{tabular}{c}
\includegraphics[scale=0.4]{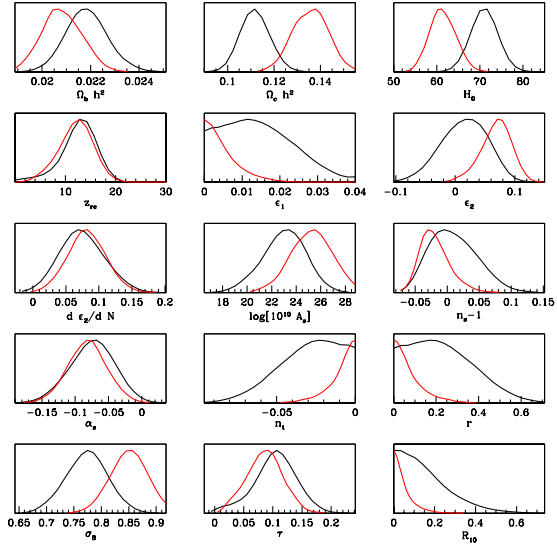}
\end{tabular}
\caption{One dimensional marginalized
likelihoods for $8$ basic parameters and $7$ derived 
parameters obtained by the GFM formula. 
We show for comparison the
constraints by using WMAP3+CMBsmall plus 2dF05
(black) or plus SDSS (red).}
\label{fig9}
\end{figure}
\begin{figure}
\begin{tabular}{ccc}
\includegraphics[scale=0.3]{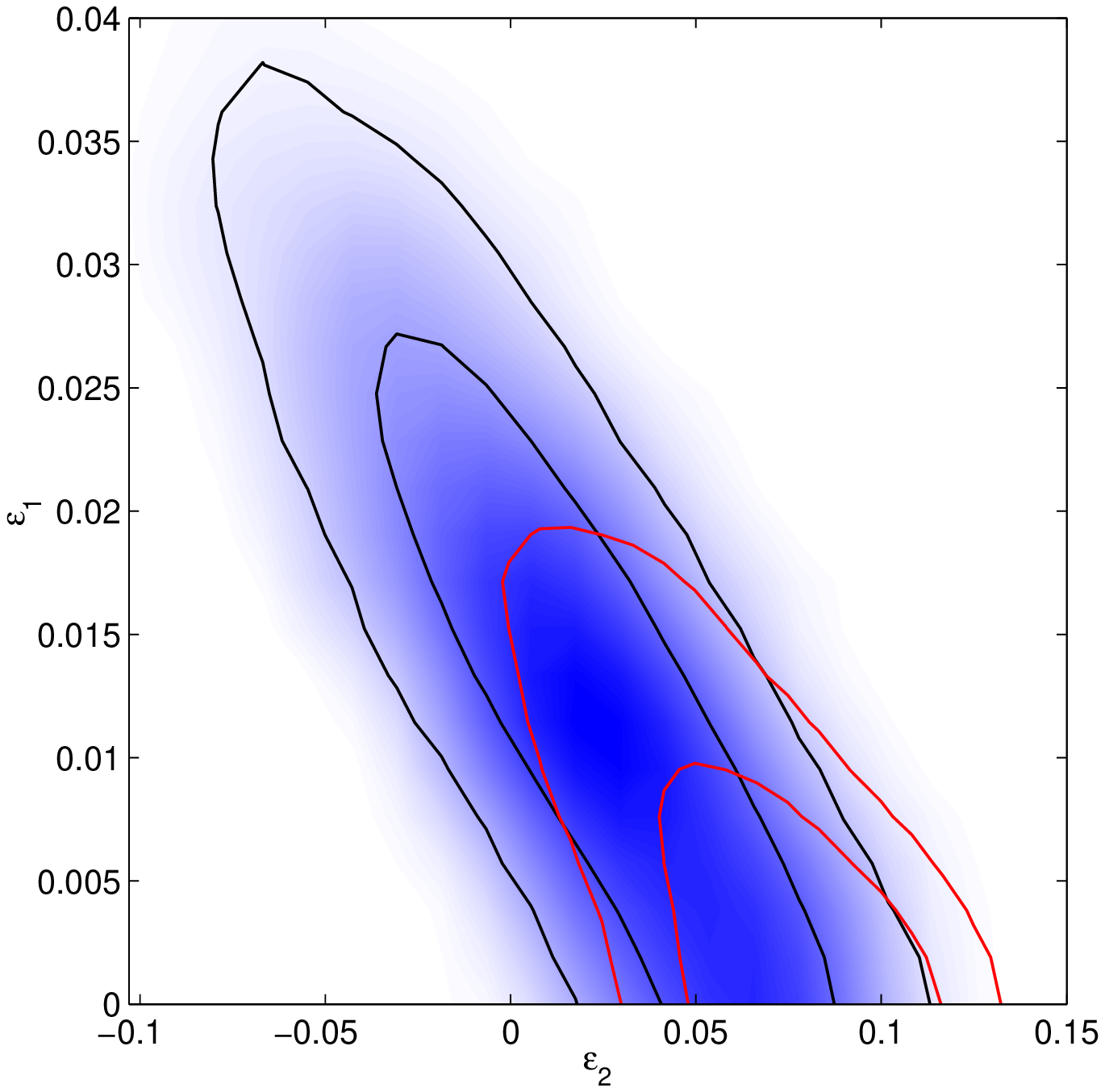}
\includegraphics[scale=0.3]{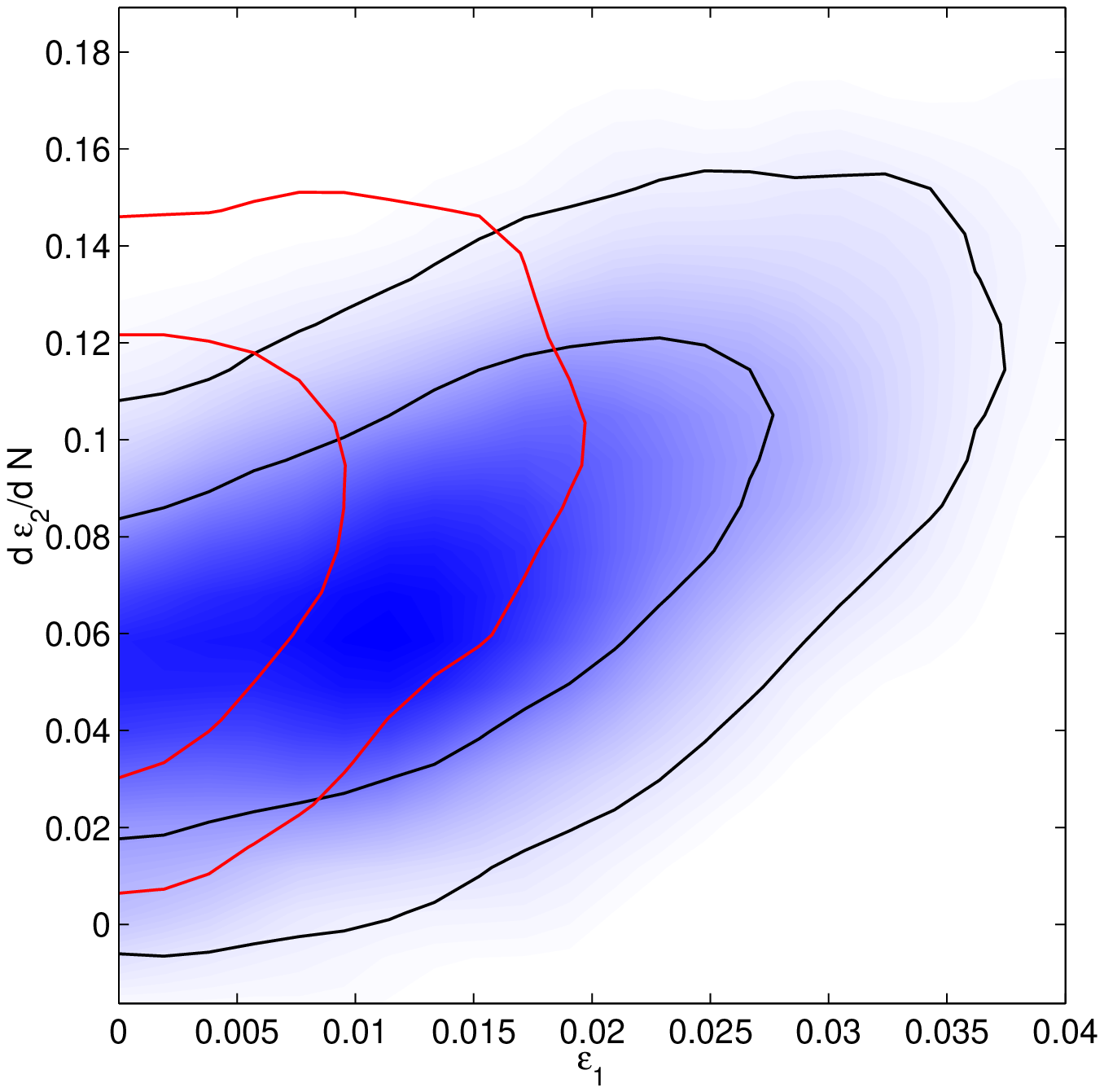}
\includegraphics[scale=0.3]{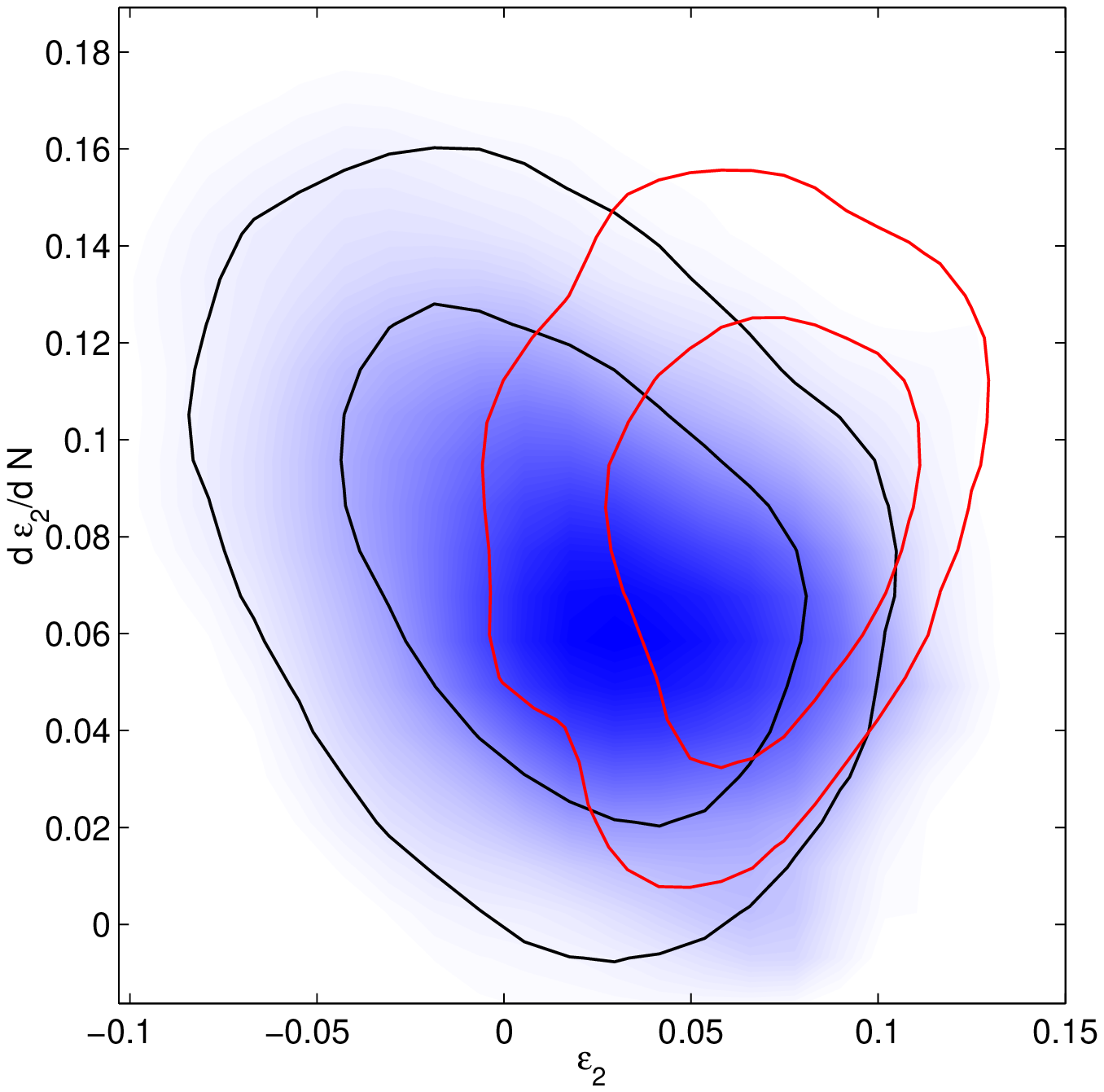}
\end{tabular}
\begin{tabular}{ccc}
\includegraphics[scale=0.3]{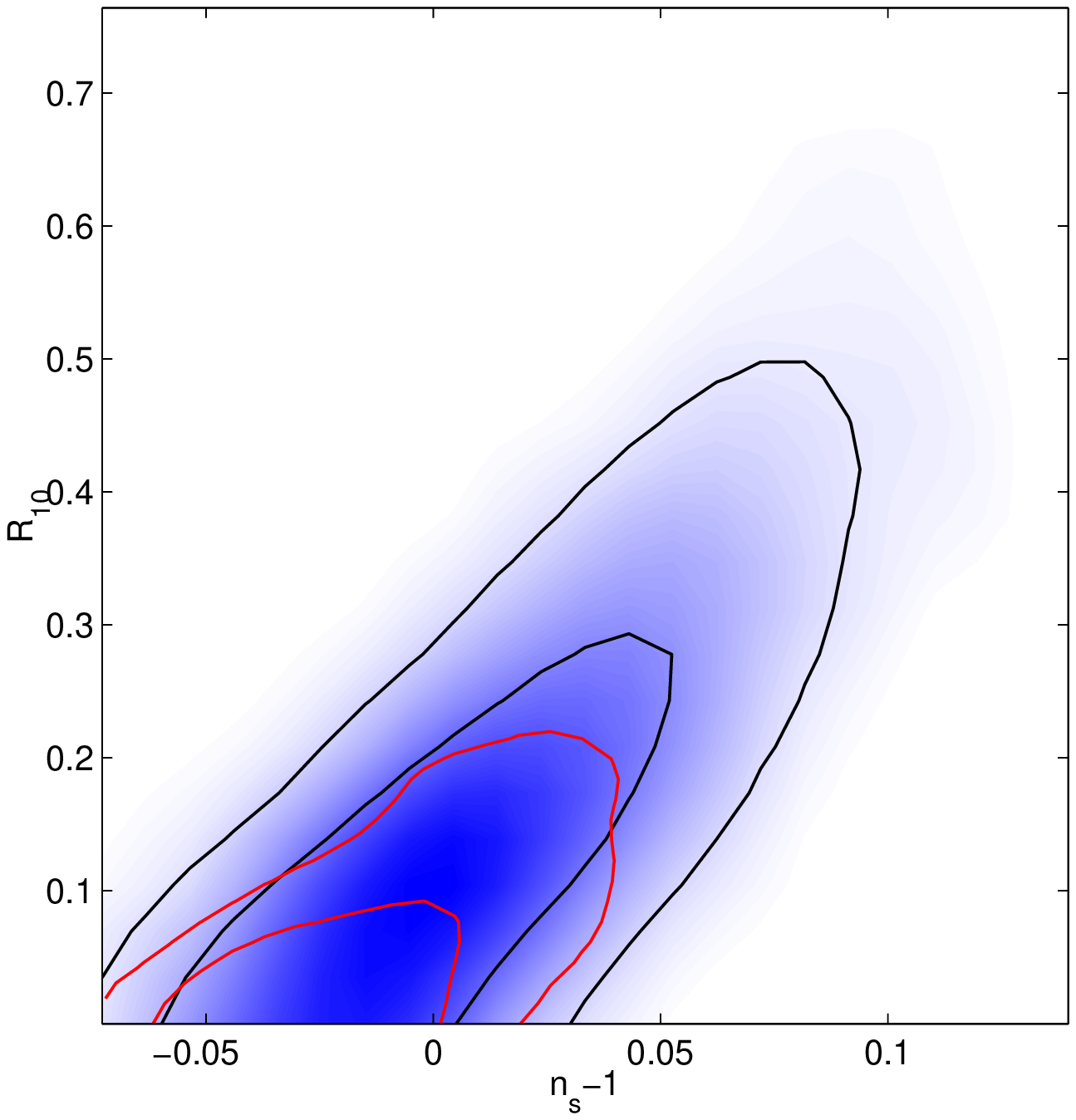}
\includegraphics[scale=0.3]{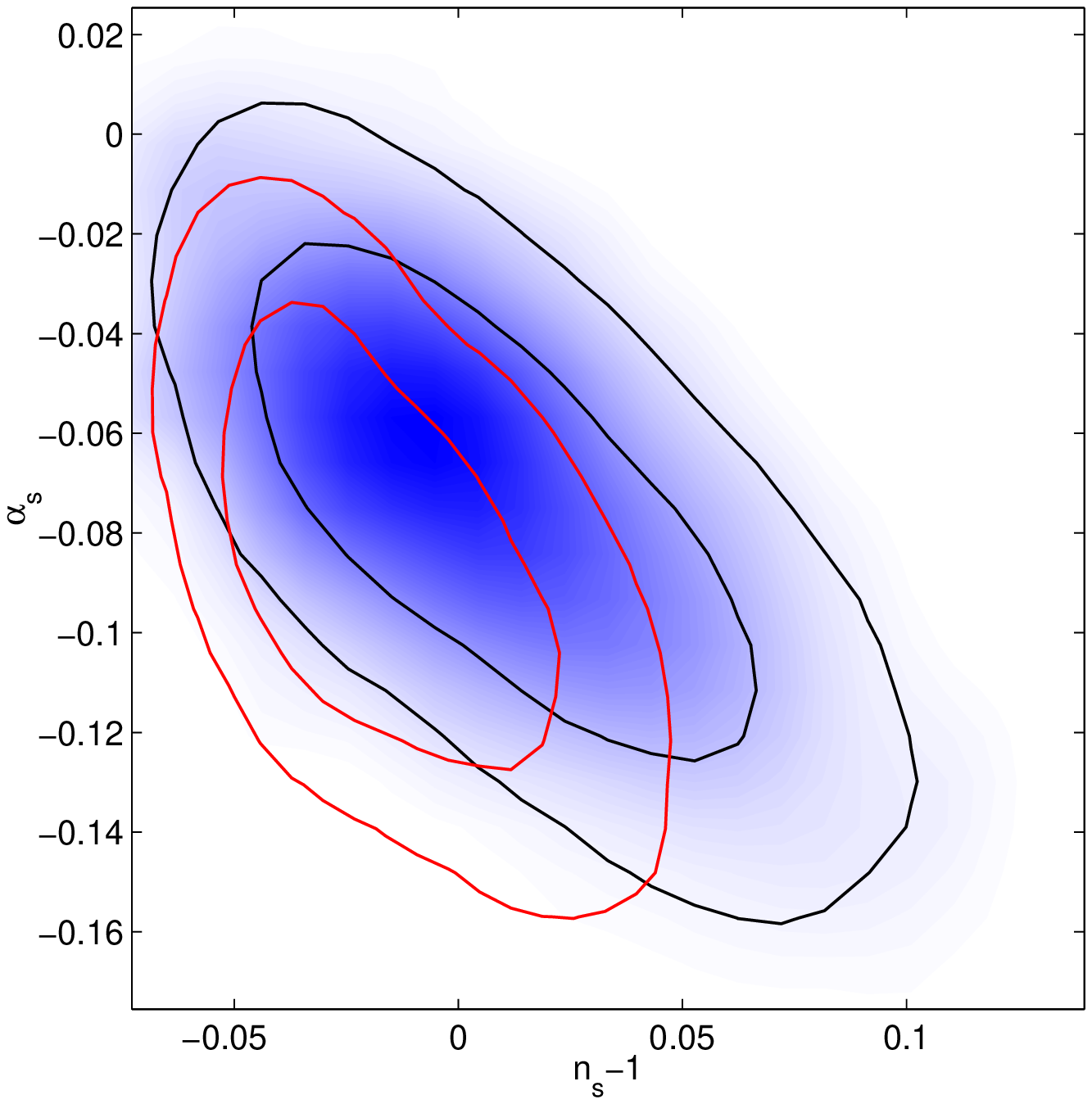}
\includegraphics[scale=0.3]{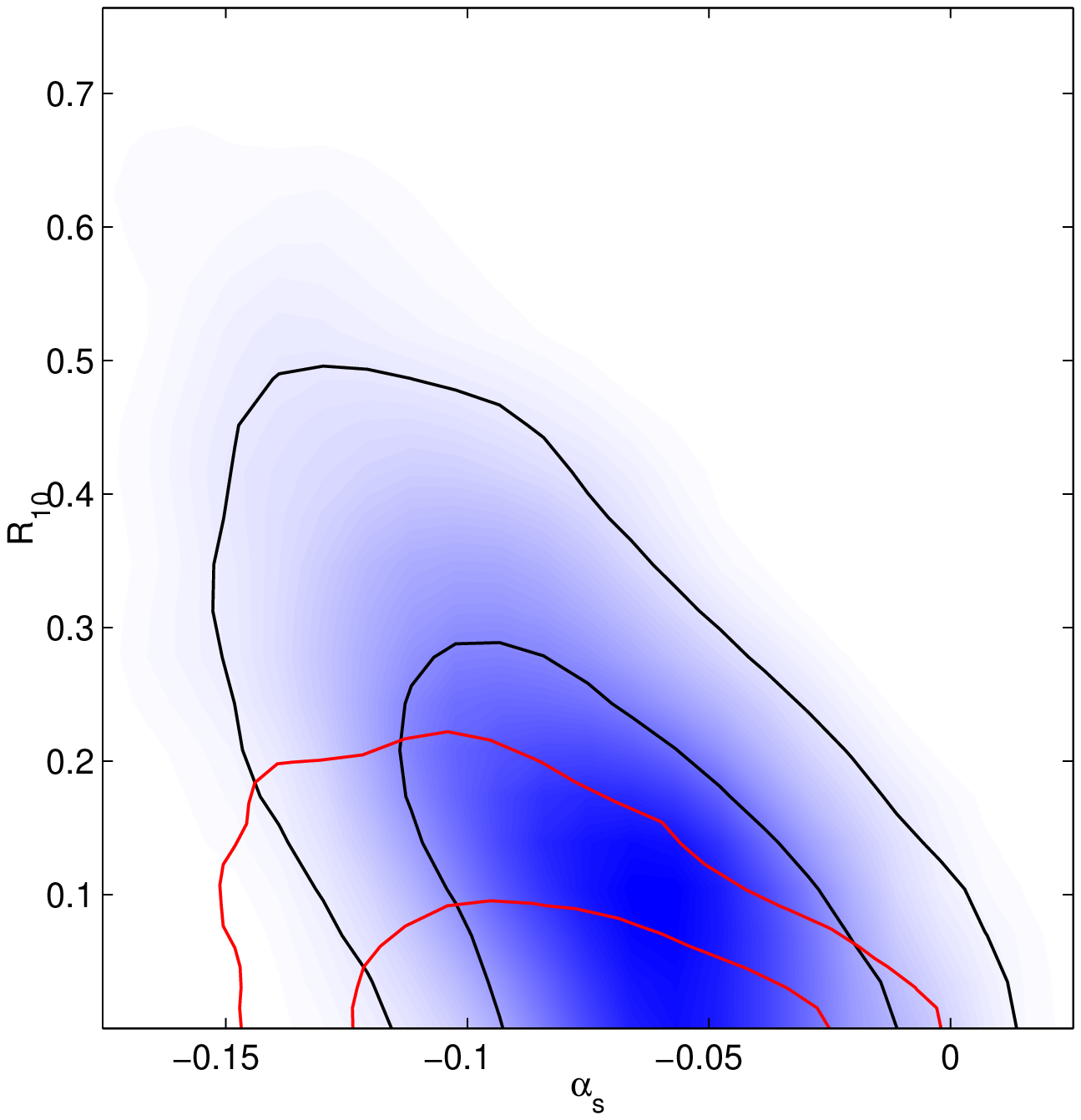}
\end{tabular}
\caption{Constraints by using SDSS (red) vs 2dF05 (black) plus
WMAP3+CMBsmall on (from left to right and top to bottom) 
($\epsilon_2 \,, \epsilon_1$), ($\epsilon_1 \,, d \epsilon_2/d N$)
($n_s - 1 \,, R_{10}$),
($n_s - 1 \,, \alpha_s$), ($\alpha_s \,, R_{10}$), ($\alpha_s \,,
\alpha_t$),
planes at $1 \sigma$ and $2 \sigma$ level.}
\label{fig10}
\end{figure}

At $1 \sigma$ we obtain the fully marginalized value of the 
running of the scalar spectral index 
$\alpha_{\rm S} = - 0.076^{+0.032}_{-0.033}$ ($-0.082^{+0.028}_{-0.028}$) 
for WMAP3+CMBsmall+2dF05 (SDSS).
The $2 \sigma$ constraint is $-0.141 < \alpha_{\rm S} < -0.013$ 
($-0.142 < \alpha_{\rm S} < -0.028$) for WMAP3+CMBsmall+2dF05 (SDSS). 
We note that the best-fit cosmological models obtained by 
2dF05 and SDSS are different, enlarging the difference between the 
two LSS data sets already seen in 
section III: SDSS prefers a very low value for $H_0$ 
(as also reported in \cite{FXY}) and the best-fit 
cosmological model with running is consistent with the one without running 
only at $2 \sigma$. The cosmological models with and without running 
for 2dF05 are consistent at $1 \sigma$: allowing running for the 
spectral indexes the constraint on $r< 0.26$ for 2dF05 is significantly 
related to $r < 0.50$ at $2 \sigma$ cl. 
 
Since $\epsilon_2 \epsilon_3$ enters only in the scalar 
running, the magnitude of the running of the tensor spectral index 
is constrained at posteriori to be 
$|\alpha_T| \sim {\cal O} (10^{-3} - 10^{-4})$, as is also clear from 
Eqs. (\ref{eqn:bt2}) and from Fig. (\ref{fig9}). 

Although our combination of data sets and our method of constraining 
inflationary parameters differ from those used 
by the WMAP team \cite{spergel}, our results are in agreement with 
those reported in \cite{spergel} for WMAP3 plus other small 
scale CMB data sets. 

\section{Conclusions}

We have analyzed the observational constraints from CMB and LSS on the 
inflationary expansion. By parametrizing the inflationary spectra as power-law 
(i.e. to first order in slow-roll parameters), we have updated 
the WMAP1 constraints of Ref. \cite{leachliddle}: the better determination 
of reionization obtained by the 3 year data release disfavour now both the 
Harrizon-Zeldovich model (with no tensor) as inflationary models 
predicting $n_s = 1$ with respect to the first year data. 
Our analysis shows that 
models with a natural exit from inflation (which stay in the 
$\epsilon_2 > 0$ region of the $(\epsilon_2 \,, \epsilon_1)$ plane according 
to \cite{dominik}) are now favoured by WMAP3. 

When running is included, we have shown how the prior on $\epsilon_3$ 
plays a crucial role in re-obtaining the WMAP team results within an 
analysis employing analytic power-spectra depending on ($H \,, 
\epsilon_i$) with $i=1-3$. Only allowing 
a broad prior on $\epsilon_3$ a fairly large negative running in 
agreement with the WMAP results is obtained. We have found that is more 
convenient to use $d \epsilon_2 / d N
= \epsilon_2 \epsilon_3$ rather than $\epsilon_3$.
The value for the running $\alpha_S$ found here implies 
$\alpha_S \ln (k/k_*)/2 > (n_S -1)$ for Fourier modes 
far from the pivot scale. 
It would be therefore interesting to  
perform a comparison of data 
with inflationary predictions at third order in the HFF. 
Note that the value of the running found here is different from zero at 
$\sim 2 \sigma$ level, and its constraints may also be affected 
by Ly$\alpha$ data \cite{viel,seljak}, which we do not consider here.

We have also studied the impact of theoretical priors 
on constraining the inflationary expansion: 
still little is known about high-energy physics and prejudices in 
interpreting cosmological data may hide interesting physics. 

We have first relaxed the hypothesis of a canonical scalar field as the 
inflaton, which leads to the consistency relation in Eq. (\ref{ttos}). 
By considering a generic Lagrangian, the constraints on the 
$(\epsilon_1 \,, \epsilon_2)$ plane degrade significantly, 
while the limits on $r$ depend on the slope of the tensor spectrum. 
This analysis shows that a KG inflaton (with sound speed $c_S=1$) 
is well inside the confidence contours of the present constraints, although the present 
data do not allow to constrain $c_S$.

We have then explored how CMB and LSS constrain inflationary models with a 
blue spectrum, a more radical departure than considering $r$ as a free 
parameter. We have 
restricted ourselves to KG field with the "wrong" sign for the 
kinetic term, in order to do not consider the tensor-to-scalar ratio as an 
independent parameter. We find that both standard inflation and NEC violating inflation 
are equally good fits to the data (the best fits for these two 
classes of models have the same $\chi^2$ 
comparing with WMAP3+CMBsmall+2dF02).
For $n_T > 0$, the constrained region on the
$(\epsilon_1 \,, \epsilon_2)$ are different from the standard case, 
leading to $0.02 \lesssim \epsilon_2 \lesssim 0.08$ with $|\epsilon_1| \lesssim 0.3$ 
at 2$\sigma$ cl. 
A simple exponential potential potential is disfavoured at 2 $\sigma$ cl 
and larger values for $r$ are allowed with respect to the standard case 
with $n_T < 0$. 


\vspace{1cm}

{\bf Acknowledgements:} 
We wish to thank Samuel Leach for help and numerous 
suggestions on the use of MCMC. 
We also thank R. Easther, W. Kinney, A. Liddle,  
H. Peiris, R. Trotta and A. Vikman for comments and discussions. 
An anonymous referee is acknowledged for her/his 
patient and critical reading of the manuscript which led to an 
improvement of this paper.
FF is partially supported by INFN ISs BO11 and PD51. 
FF thanks the Galileo Galilei Institute for Theoretical Physics for the
hospitality and
the INFN for partial support during the completion of this work
Part of the results presented here have been 
obtained on CINECA Linux cluster under the agreement INAF/CINECA. 
We acknowledge the use of the package {\sc cosmomc} and the use of 
the Legacy Archive for Microwave Background Data Analysis (LAMBDA). 
Support for LAMBDA is provided by the NASA Office of Space Science.


\end{document}